\documentclass[a4paper,twocolumn,accepted=2025-11-24]{quantumarticle}
\pdfoutput=1
\usepackage{amssymb,amsmath,amsthm,bbm,bbold}
\usepackage{amsfonts}
\usepackage{graphicx}
\usepackage{enumitem}
\usepackage[arrow, matrix, curve]{xy}
\usepackage[numbers,sort&compress]{natbib}
\usepackage{mathtools}
\usepackage{tcolorbox}

\usepackage[normalem]{ulem}
\usepackage{color}
\usepackage{xkeyval, etoolbox, geometry, fancyhdr, tikz, ltxgrid, hyperref, ltxcmds}
\usepackage{xcolor}

\def\RR{\mathbbm{R}}
\def\CC{\mathbbm{C}}

\newcommand{\bd}[1]{\boldsymbol{#1}}
\newcommand{\wb}{\bd{w}}

\usepackage[makeroom]{cancel}

\newcommand*\xbar[1]{\hbox{\vbox{
       \hrule height 0.6pt 
       \kern0.3ex
       \hbox{%
         \kern-0.2em
         \ensuremath{#1}%
         \kern 0.0em
         }}}}

\newcommand*\xxbar[1]{\hbox{\vbox{
       \hrule height 0.6pt 
       \kern0.3ex
       \hbox{%
         \kern-0.0em
         \ensuremath{#1}%
         \kern 0.0em
         }}}}
\usepackage{accents}

\newcommand{\Ebw}{\,\xxbar{\mathcal{E}}^N_{S,M}\!(\wb)}
\newcommand{\ebw}{\,\xxbar{\mathcal{L}}^1_{N,S}(\wb)}

\definecolor{ao}{rgb}{0.0, 0.5, 0.0}

\newcommand{\EbwS}{\,\scriptsize{\xbar{\mathcal{E}}}\normalsize^N_{S,M}\hspace{-0.7mm}(\wb)}

\newcommand{\Tr}{\mathrm{Tr}}

\newcommand{\bra}[1]{\mbox{$\langle #1 |$}}
\newcommand{\ket}[1]{\mbox{$| #1 \rangle$}}

\begin{document}
\title{Solving one-body ensemble N-representability problems with spin}

\author{Julia Liebert}
\affiliation{Department of Physics, Arnold Sommerfeld Center for Theoretical Physics,
Ludwig-Maximilians-Universit\"at M\"unchen, Theresienstrasse 37, 80333 M\" unchen, Germany}
\affiliation{Munich Center for Quantum Science and Technology (MCQST), Schellingstrasse 4, 80799 M\"unchen, Germany}

\author{Federico Castillo}
\affiliation{Departamento de Matem\'aticas, Pontificia Universidad Cat\'olica de Chile, Santiago, Chile}

\author{Jean-Philippe Labb\'e}
\affiliation{\'Ecole de Technologie Sup\'erieure, 1111 rue Notre-Dame Ouest, Montr\'eal (Qc) H3C 6M8, Canada}

\author{Tomasz Maciazek}
\affiliation{School of Mathematics, University of Bristol, Fry Building, Woodland Road, Bristol, BS8 1UG, United Kingdom}

\author{Christian Schilling}
\email{c.schilling@lmu.de}
\affiliation{Department of Physics, Arnold Sommerfeld Center for Theoretical Physics,
Ludwig-Maximilians-Universit\"at M\"unchen, Theresienstrasse 37, 80333 M\" unchen, Germany}
\affiliation{Munich Center for Quantum Science and Technology (MCQST), Schellingstrasse 4, 80799 M\"unchen, Germany}


\begin{abstract}
The Pauli exclusion principle is fundamental to understanding electronic quantum systems, imposing constraints on the expected occupancies $n_i$ of orbitals $\varphi_i$, such that $0 \leq n_i \leq 2$. In this work, we refine the underlying one-body $N$-representability problem by incorporating spin symmetries and a potential degree of mixedness $\wb$ of the $N$-electron quantum state. Employing basic tools from representation theory, convex analysis, and discrete geometry, we derive a comprehensive solution to this problem. Specifically, we demonstrate that the set of admissible orbital one-body reduced density matrices is fully characterized by linear spectral constraints on the natural orbital occupation numbers, defining a convex polytope $\Sigma_{N,S}(\wb) \subset [0,2]^d$. These constraints are independent of the magnetization $M$ and the number $d$ of orbitals, while their dependence on $N$ and the total spin $S$ is linear, and we thus calculate them for arbitrary system sizes and spin quantum numbers. Our results provide a crucial missing cornerstone for ensemble density (matrix) functional theory.
\end{abstract}

\maketitle

\section{Introduction}\label{sec:intro}
The original Pauli exclusion principle has profoundly shaped our understanding of $N$-electron quantum systems, as it imposes the fundamental restriction that each orbital can accommodate at most one spin-up and one spin-down electron. This principle lies at the core of the Aufbau principle, which explains the structure of the periodic table of elements. When incorporating the probabilistic nature of quantum mechanics and the influence of electron correlation, the Pauli exclusion principle can be reformulated as a universal kinematic constraint on the orbital one-particle reduced density matrix (1RDM) $\gamma_l$: for any $N$-electron quantum state, the vector of orbital occupancies $\bra{\varphi_i} \gamma_l \ket{\varphi_i}$ of $d$ reference orbitals $\varphi_i$ is confined to the Pauli hypercube $[0, 2]^d$. This geometric interpretation corresponds to the solution of the so-called one-body ensemble $N$-representability problem, which seeks to characterize the set of orbital 1RDMs $\gamma_l$ that can be derived from some $N$-electron density matrix \cite{C63, Coleman-book}.

Despite its broad relevance, the Pauli exclusion principle ignores two critical features of realistic $N$-electron quantum states, particularly in the context of quantum chemistry. First, these states are often characterized by definite spin quantum numbers, reflecting the spin symmetries of the underlying Hamiltonian. Furthermore, electron spin plays a fundamental role in various physical phenomena, such as magnetism and the quantum Hall effect \cite{vKDP80, vK86, FNTW98, MR03}, and its incorporation in numerical methods is essential for achieving high predictive accuracy. Second, due to thermal effects or entanglement arising from interactions with the environment, $N$-electron states frequently exhibit a certain degree of mixedness. It is therefore the primary goal of this work to introduce and solve a refined version of the one-body ensemble $N$-representability problem that accounts for these two crucial aspects. To achieve this, we employ basic mathematical tools to construct a comprehensive solution. This approach leads to a generalized exclusion principle that, from a geometric perspective, identifies a subpolytope of the Pauli hypercube, $\Sigma_{N,S}(\wb) \subset [0,2]^d$, representing the admissible natural orbital occupation numbers.

From a practical point of view, our work is motivated by the growing interest in reduced density matrix methods for targeting excited states using a distinctive ensemble variational principle \cite{GOK88a,LHS24}. Specifically, we demonstrate that solving the spin-symmetry-adapted ensemble $N$-representability problem provides a compact and rigorous characterization of the yet-unknown domain of universal interaction functionals in the rapidly evolving field of ensemble density functional theory (EDFT) \cite{Theophilou79, GOK88c, TG95, YPBU17, Fromager2020-DD, Loos2020-EDFA, Cernatic21, GK21, Yang21, GKGGP23,GL23, SKCPJB24, CPSF24} and ensemble one-particle reduced density matrix functional theory ($\wb$-RDMFT) \cite{SP21, LCLS21, LS23-njp, LS23-sp}. In this context, our work addresses a critical gap, providing a foundational cornerstone for the advancement of these ensemble-based methodologies.

This paper is organized as follows. In Sec.~\ref{sec:concepts}, we introduce the notation and foundational concepts needed to address spin symmetries at both the one- and $N$-particle levels (Sec.~\ref{sec:notation}). We then formally define the orbital one-body $\bd{w}$-ensemble $(N, S, M)$-representability problem in Sec.~\ref{sec:science}. In Sec.~\ref{sec:QMP}, we develop a general solution to this problem for fermions, highlighting its key properties. Sec.~\ref{sec:examples} provides explicit computations and illustrative examples of the derived constraints. Finally, in Sec.~\ref{sec:appl}, we demonstrate three applications of these new constraints in the context of reduced density matrix methods.

\section{Key concepts and scientific problem \label{sec:concepts}}

In Sec.~\ref{sec:notation}, we introduce some notation and basic concepts such as spin symmetries and reduced density matrices. This will then allow us to formally introduce the symmetry-adapted orbital one-body $\bd w$-ensemble $N$-representability problem in Sec.~\ref{sec:science}, which is the scientific problem addressed and comprehensively solved in this work.

\subsection{Notation\label{sec:notation}}

We consider non-relativistic spin-$1/2$ fermions whose one-particle Hilbert space $\mathcal{H}_1$ exhibits the tensor product structure
\begin{equation}\label{eq:H1}
\mathcal{H}_1=\mathcal{H}_1^{(l)}\otimes\mathcal{H}_1^{(s)}\cong \CC^d\otimes\CC^2\,,
\end{equation}
where $\mathcal{H}_1^{(l)}$ denotes the orbital component of $\mathcal{H}_1$ with dimension $\mathrm{dim}(\mathcal{H}_1^{(l)})=d<\infty$, and $\mathcal{H}_1^{(s)}$ describes the spin degrees of freedom.
The Hilbert space of $N$-fermions, $\mathcal{H}_N$, follows as the $N$-fold wedge product of $\mathcal{H}_1$, that is $\mathcal{H}_N = \wedge^N\mathcal{H}_1$.
Due to the relevance of spin symmetries in physics and chemistry as explained in the introduction, we consider $N$-fermion states with well-defined total spin $S$ and magnetization $M$. The magnetization $M$ follows in general as the expectation value of the $S_z$ operator,
\begin{equation}
S_z = \frac{1}{2}\sum_{i=1}^d\left(f_{i\alpha}^\dagger f_{i\alpha}- f_{i\beta}^\dagger f_{i\beta}\right)\,,
\end{equation}
where $f_{i\sigma}^\dagger, f_{i\sigma}$ denote the fermionic creation/annihilation operators for a fermion in a spatial orbital $i$ with spin $\sigma=\alpha, \beta$\footnote{We resort in our work to a notation for the two spin states used commonly in quantum chemistry, also since the symbols $\uparrow/\downarrow$ will be used to indicate vectors with increasingly/decreasingly ordered entries.} and we set $\hbar\equiv1$.
Moreover, the total spin quantum number $S$ is determined by the so-called Casimir operator $\bd S^2$ of the Lie group $SU(2)$,
\begin{equation}
\bd S^2 =  \sum_{i,j=1}^dS_i^zS^z_j+\frac{1}{2}\left(S^+_iS^-_j + S^-_iS^+_j\right)\,,
\end{equation}
where $S_i^+= f_{i\alpha}^\dagger f_{i\beta}, S_i^-=f_{i\beta}^\dagger f_{i\alpha}$ denote the spin raising and lowering operators.
The operators $S_z, S_{\pm}$ constitute the generators of the group $SU(2)$ and commute with spin-independent Hamiltonians as, for instance, the non-relativistic electronic structure Hamiltonian. Which such applications in mind, it is then advantageous to exploit the Peter-Weyl decomposition of the $N$-fermion Hilbert space $\mathcal{H}_N$ into symmetry sectors labelled by the quantum numbers $S, M$ according to (see, e.g., \cite{Hall15, Miller72})
\begin{equation}
\mathcal{H}_N = \bigoplus_{S=S_{\mathrm{min}}}^{N/2}\bigoplus_{M=-S}^S\mathcal{H}_N^{(S,M)}\,,
\end{equation}
where $S_{\mathrm{min}}=0$ for $N$ even and $S_{\mathrm{min}}=1/2$ for $N$ odd.

The set of density operators corresponding to pure quantum states with spin quantum numbers $S,M$ follows as
\begin{eqnarray}\label{eq:PNSM_def}
\mathcal{P}^N_{S,M} &\equiv& \Big\{\Gamma:\mathcal{H}_N^{(S,M)}\to\mathcal{H}_N^{(S,M)}\,|\,\Gamma\geq 0, \nonumber \\ &&\hspace{0.5cm}\Tr_N[\Gamma]=1,\Gamma=\Gamma^2\Big\}\,.
\end{eqnarray}
Here, it is taken into account that for finite-dimensional Hilbert spaces positivity $\Gamma\geq 0$ implies hermiticity, $\Gamma=\Gamma^\dagger$, and the condition $\Gamma=\Gamma^2$ together with the normalization implies that $\Gamma=\ket{\Psi}\!\bra{\Psi}$ for some suitable $\ket{\Psi} \in \mathcal{H}_N^{(S,M)}$. What is more, the elements of the set $\mathcal{P}^N_{S,M}$ constitute the extremal points of the convex and compact set $\mathcal{E}^N_{S,M}$ of all ensemble $N$-fermion quantum states,
\begin{equation}\label{eq:ENSM_def}
\mathcal{E}^N_{S,M}=\mathrm{conv}\left(\mathcal{P}^N_{S,M}\right)\,,
\end{equation}
where here and in the following $\mathrm{conv}(\cdot)$ always denotes the convex hull operation. The one-particle reduced density matrix (1RDM) $\gamma$ of a quantum state $\Gamma\in\mathcal{E}^N_{S,M}$ is obtained by tracing out all except one fermion according to $\gamma=N\Tr_{N-1}[\Gamma]$. Conservation of the total magnetization $M$ implies that the 1RDM $\gamma$ is block-diagonal with respect to the spin states $\ket{\alpha}, \ket{\beta}$ and the only non-vanishing blocks in $\gamma$ are the $\alpha, \beta$-blocks, $\gamma_{\alpha\alpha}, \gamma_{\beta\beta}$, i.e.,
\begin{equation}\label{eq:gblockdiag}
  [S_z,\Gamma]=0 \quad \Rightarrow \quad\gamma = \gamma_{\alpha\alpha} \oplus \gamma_{\beta\beta}\,.
\end{equation}
Furthermore, tracing out the spin degree of freedom yields the so-called orbital (or spin-traced) 1RDM
\begin{equation}
\gamma_l \equiv\mathrm{Tr}_{\mathcal{H}_1^{(s)}}[\gamma]= \gamma_{\alpha\alpha}+\gamma_{\beta\beta}\,.
\end{equation}
Accordingly, the orbital 1RDM $\gamma_l$ is obtained from $\Gamma$ by applying the map $\mu$
\begin{equation}\label{eq:map}
\mu = \mathrm{Tr}_{\mathcal{H}_1^{(s)}}\circ N\Tr_{N-1}\,,
\end{equation}
a composition of two partial traces. Since partial traces are linear, $\mu$ is a linear map on the space of (density) operators, a property that will be crucial for our work.
The matrix elements of the orbital 1RDM in a given basis of $\mathcal{H}_1^{(l)}$,
\begin{equation}\label{eq:1RDM}
\bra{i} \gamma_l \ket{j}\equiv (\gamma_l)_{ij} = \mathrm{Tr}_N[E_{ij}\Gamma]\,,
\end{equation}
are nothing else than the expectation values of the $U(d)$ generators \cite{Moshinsky69, RCR12}
\begin{equation}\label{eq:Eij}
E_{ij} = f_{j  \alpha}^\dagger f_{i\alpha} + f_{j\beta}^\dagger f_{i\beta}\,.
\end{equation}
These generators will be essential for deriving the solution to the orbital one-body $\bd w$-ensemble $(N, S,M)$-representability problem introduced in the next section.

\subsection{Orbital one-body $\bd w$-ensemble $(N,S,M)$-representability problem \label{sec:science}}


The set of orbital 1RDMs $\gamma_l$, which are compatible to a pure $N$-fermion state $\Gamma\in\mathcal{P}^N_{S,M}$ is denoted by
\begin{equation}\label{eq:L1N}
\mathcal{L}_{N,S}^1 = \left\{\gamma_l\,|\, \gamma_l = \mu(\Gamma)\text{ for some }\Gamma\in \mathcal{P}^N_{S,M}  \right\}\,.
\end{equation}
Since the orbital 1RDM is spin-traced, this and various other sets of orbital 1RDMs introduced throughout this work are independent of the magnetic quantum number $M$.
We therefore omit indices $M$ whenever possible.

Since the map $\mu(\cdot)$ is linear, the image of the set $\mathcal{E}^N_{S,M}$ under the map $\mu(\cdot)$ \eqref{eq:map} is equal to the convex hull of $\mathcal{L}_{N,S}^1$, such that
\begin{align}\label{eq:Lbar1N}
\xxbar{\mathcal{L}}_{N,S}^1&\equiv \mu\left(\mathcal{E}^N_{S,M}\right)= \mathrm{conv}\left(\mathcal{L}_{N,S}^1 \right)\,.
\end{align}
An orbital 1RDMs $\gamma_l$ is called \emph{pure state $(N,S,M)$-representable}
if and only if $\gamma_l \in \mathcal{L}_{N,S}^1$.
Similarly, $\gamma_l$ is called \emph{ensemble $(N,S,M)$-representable} if and only if $\gamma_l \in \xxbar{\mathcal{L}}_{N,S}^1$. 
The sets of pure/ensemble $(N,S,M)$-representable orbital 1RDMs are invariant under unitary transformations $u_l$ on $\mathcal{H}_1^{(l)}$, i.e.,
\begin{equation}\label{eq:unitinv}
\gamma\in \mathcal{L}_{N,S}^1\quad\Rightarrow\quad u_l\gamma u_l^\dagger\in \mathcal{L}_{N,S}^1
\end{equation}
and analogously for $\xxbar{\mathcal{L}}_{N,S}^1$. Therefore, for any values $N,S,M$ and $d$, both sets are fully characterized by purely \emph{spectral} constraints,   conditions on the vector $\bd\lambda$ of eigenvalues of $\gamma_l$, also called \emph{natural orbital occupation numbers}.
A formal mathematical procedure for deriving the spectral characterization of the set $\mathcal{L}_{N,S}^1$ has been provided by Klyachko and Altunbulak in Refs.~\cite{KL06, AK08}. Yet, due to its mathematical complexity, this problem could be solved only for artificially small values $N,d$. Quite in contrast, by using tools from convex analysis, a complete spectral characterization of $\xxbar{\mathcal{L}}_{N,S}^1$ could be derived recently in Ref.~\cite{LMS23} for arbitrary system sizes $N,d$. In analogy to the spin-adapted generalized Pauli constraints derived by Klyachko and Altunbulak in Refs.~\cite{KL06, AK08}, these convex relaxed spin-adapted exclusion principle constraints take the form of linear inequalities resulting in a convex spectral polytope, a spin-dependent subset of the Pauli hypercube $[0,2]^d$.

Moreover, we like to stress that the pure state setting of Refs.~\cite{KL06, AK08} means to restrict to $N$-fermion density matrices $\Gamma$ with specific fixed spectrum $\mathrm{spec}^\downarrow(\Gamma) = \bd w_0 \equiv (1, 0, \ldots)$. Therefore and due to the physical relevance of mixed states, one might be also interested in the solution of the one-body $N$-representability problem for a generic choice of the fixed spectrum $\mathrm{spec}^\downarrow(\Gamma) = \bd w=\bd w^{\downarrow}$, where $w_1\geq w_2\geq ...\geq w_D$. 
This more general one-body $N$-representability problem contains indeed Klyachko's pure state $N$-representability problem as a special instance, namely for $\bd w = \bd w_0$. This also implies that solving the more general problem with generic $\bd w$ will be more complicated than the one for pure states.

To this end, we first introduce the respective sets of $N$-fermion states and orbital 1RDMs. The set of $N$-fermion states $\Gamma\in \mathcal{E}^N_{S,M}$  with fixed spectrum $\bd w$ is denoted by
\begin{equation}\label{def:ENSMw}
\mathcal{E}^N_{S,M}(\bd w) \equiv \left\{\Gamma\in \mathcal{E}^N_{S,M}\,|\, \mathrm{spec}^\downarrow(\Gamma) = \bd w\right\}\,.
\end{equation}
The set $\mathcal{E}^N_{S,M}(\bd w)$ is in general not convex, as the convex combination of two states with fixed spectrum $\bd w$ usually yields a state with a different spectrum. Therefore, we also introduce its convex hull,
\begin{equation}\label{def:ENSMwrelax}
\Ebw \equiv \mathrm{conv}\left(\mathcal{E}^N_{S,M}(\bd w)\right)\,.
\end{equation}
A basic mathematical result (see, e.g., Ref.~\cite{LCLS21}) provides a simple characterization of such
convex hulls as the one in Eq.~\eqref{def:ENSMwrelax}: A density matrix $\Gamma$ belongs to the set $\Ebw$ if and only if its spectrum is majorized by $\wb$, $\mathrm{spec}^\downarrow(\Gamma) \prec \wb$. In general, we say that a vector $\bd x \equiv (x_1,\ldots,x_D)$ majorizes a vector $\bd y \equiv (y_1,\ldots,y_D)$, $\bd y\prec \bd x$, if and only if for all $k=1,2,\ldots, D$ the following condition holds
\begin{equation}\label{eq:majoriz}
y_1^\downarrow+\ldots + y_k^\downarrow \leq x_1^\downarrow+\ldots + x_k^\downarrow  \,,
\end{equation}
with equality for $k=D$ and the superscript $\downarrow$ indicates that the entries in a vector are rearranged decreasingly, i.e., the entries of $\bd x^\downarrow$ satisfy $x_1^\downarrow\geq x_2^\downarrow\geq ...\geq x_D^\downarrow$. The majorization $\prec$ defines a (pre)order on the space of vectors. Moreover, when applied to the spectra of density matrices it compares them in terms of their degree of mixedness. In particular, $\mathrm{spec}(\Gamma) \prec \mathrm{spec}(\Gamma')$ implies $S(\Gamma)\geq S(\Gamma')$, where $S(\Gamma)\equiv - \mbox{Tr}[\Gamma \log \Gamma]$ is the von Neumann entropy \cite{Marshall}.

Applying the map $\mu(\cdot)$ in Eq.~\eqref{eq:map} to the $N$-fermion states in $\mathcal{E}^N_{S,M}(\bd w), \Ebw$ yields the following sets of admissible orbital 1RDMs,
\begin{align}\label{eq:L1Nw}
\mathcal{L}_{N,S}^1 (\bd w)&\equiv \mu\left(\mathcal{E}^N_{S,M}(\bd w)\right)\,,\\ \nonumber
\ebw&\equiv\mu\left(\Ebw\right)\,.
\end{align}
\begin{figure}[tb]
\centering
\frame{\includegraphics[width=0.8\linewidth]{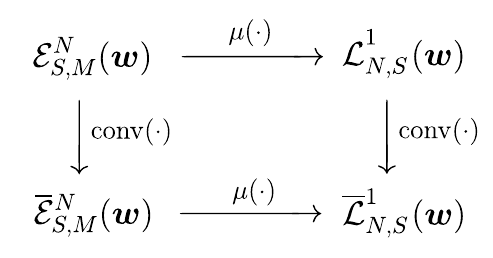}}
\caption{Commutative diagram illustrating the relation among the two non-convex sets $\mathcal{E}^N_{S,M}(\bd w), \mathcal{L}^1_{N,S,M}(\bd w)$ and their convex hulls $\xxbar{\mathcal{E}}^N_{S,M}(\bd w),\xxbar{\mathcal{L}}^1_{N,S,M}(\bd w)$ via the map $\mu(\cdot)$. \label{fig:mu-conv-map}}
\end{figure}
As it is explained and illustrated in Fig.~\ref{fig:mu-conv-map}, the set $\ebw$ is the convex hull of $\mathcal{L}_{N,S}^1(\bd w)$, in analogy to Eq.~\eqref{eq:Lbar1N} for the specific case $\bd w = \bd w_0$,
\begin{equation}\label{eq:Lbar1Nw}
   \ebw =   \mathrm{conv}\left(\mathcal{L}_{N,S}^1(\bd w)\right)\,.
\end{equation}
The relations among the four sets $\mathcal{E}^N_{S,M}(\bd w), \Ebw, \mathcal{L}_{N,S}^1, \ebw$ are illustrated in Fig.~\ref{fig:mu-conv-map}.
\begin{figure*}[htb]
\centering
\frame{\includegraphics[width=\linewidth]{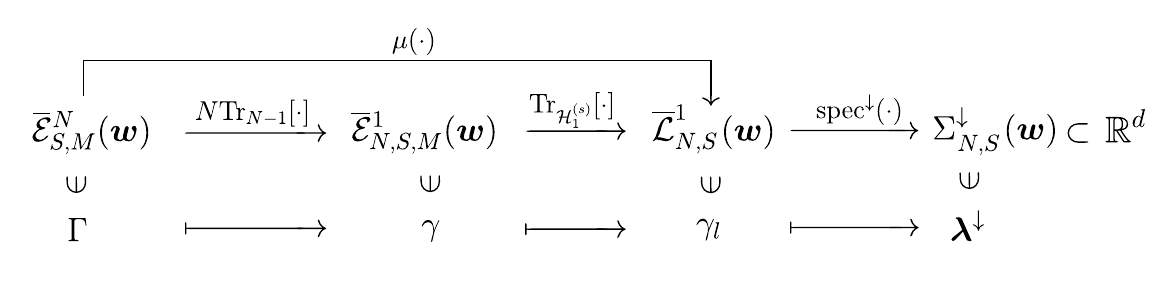}}
\caption{Illustration of the relations between the most relevant sets of (reduced) density matrices and spectra introduced in Sec.~\ref{sec:science}. For the sake of clarity, we also present as an intermediate level the corresponding set $\xxbar{\mathcal{E}}^1_{N,S, M}(\wb)$ of full  1RDMs $\gamma$. Accordingly, the map $\mu(\cdot)$ introduced in Eq.~\eqref{eq:map} maps an $N$-fermion density matrix $\Gamma\in\mathcal{E}^N_{S,M}$ to its orbital 1RDM $\gamma_l$. The solution to the relaxed orbital one-body $\bd w$-ensemble $N$-representability problem characterizes the spectral set $\Sigma_{N,S}^{\downarrow}(\bd w)$.
\label{fig:map-mu}}
\end{figure*}


We are now in a position to present in concise terms the scientific problem introduced and addressed by our work:
\begin{tcolorbox}[colback=ao!20, colframe=ao, left=2pt,right=2pt, title= Scientific Problem: Spin-symmetry adapted orbital one-body $\boldsymbol{w}$-ensemble $N$-representability problem]
For given weights $\wb$ and values $N,d,S,M$, determine necessary and sufficient conditions that describe whether a given orbital 1RDM $\gamma_l$ on $\mathcal{H}_1^{(l)} \cong\CC^d$ is representing an $N$-fermion density matrix with spin quantum numbers $S,M$ and spectrum at least as mixed as $\wb$, i.e., $ \mathrm{spec}^\downarrow(\Gamma) \prec \wb$ (recall Eq.~\eqref{eq:majoriz}). By recalling definitions \eqref{eq:map},\eqref{def:ENSMw},\eqref{def:ENSMwrelax} and Fig.~\ref{fig:map-mu}, this means to find an efficient characterization of the convex set
\begin{eqnarray}
\ebw &\equiv& \mu(\Ebw) \nonumber \\
&=& \mu\left(\mathrm{conv}\left(\mathcal{E}^N_{S,M}(\bd w)\right)\right) \,.
\end{eqnarray}
We refer to $\gamma_l \in \ebw$ as being \emph{(relaxed) $\wb$-ensemble $($N,S,M$)$-representable}.
\end{tcolorbox}

We provide a comprehensive solution to this relaxed orbital one-body spin-adapted $N$-representability problem in the subsequent sections. In order to complete the present section a couple of comments are in order concerning (i) the form of the anticipated solution, (ii) its physical relevance and (iii) the connection of our work to previous ones on related topics:
\begin{enumerate}[label=(\roman*)]
\item First, as already stressed at the beginning of Sec.~\ref{sec:concepts} and as our derivation below will make explicit, various sets of orbital 1RDMs defined throughout this work (such as those in Eqs.~\eqref{eq:L1N},\eqref{eq:Lbar1N},\eqref{eq:L1Nw}) are independent of the magnetic quantum number $M$.
    Moreover, as $\ebw $ is invariant under unitary transformations on the orbital one-particle Hilbert space $\mathcal{H}_1^{(l)}$, in analogy to Eq.~\eqref{eq:unitinv}, the solution to the relaxed $(S,N,M)$-representability problem is fully characterized by the spectral set
\begin{align}
\Sigma_{N,S}(\bd w) \equiv \Big\{\bd\lambda\,|\,&\exists\pi\in\mathcal{S}_d, \gamma_l\in\ebw :\nonumber\\
& \bd\lambda = \pi(\mathrm{spec}^\downarrow(\gamma_l))\Big\}\,,
\end{align}
where $\mathcal{S}_d$ denotes the symmetric group of degree $d$. The symmetric group $\mathcal{S}_d$ acts on a vector $\bd\lambda^\downarrow = (\mathrm{spec}^\downarrow(\gamma_l))\in \RR^d$ by permutations of its $d$ entries.
Moreover, it is sufficient to consider only the set $\Sigma_{N,S}^\downarrow(\bd w)$ of decreasingly ordered natural orbital occupation number vectors $\bd\lambda$. Both sets $\Sigma_{N,S}(\bd w), \Sigma_{N,S}^\downarrow(\bd w) $ are at most $(d-1)$-dimensional due to the normalization condition on $\bd \lambda$, $\sum_{j=1}^d\lambda_j=N$. We refer the reader to Fig.~\ref{fig:map-mu} for a schematic illustration and summary of the relations between the different sets of density matrices and spectral sets introduced in this section. Finally, we anticipate that the sought-after sets $\Sigma_{N,S}(\bd w), \Sigma_{N,S}^\downarrow(\bd w)$ are taking the form of convex polytopes, i.e., the solution to the scientific problem is given by a finite family of linear conditions on the natural orbital occupation numbers. These $\wb$-ensemble constraints are found to be effectively form-independent of $N,S$ and $d$. In particular, a remarkable hierarchy is discovered by referring to the number $r$ of non-vanishing entries in $\wb$. To be more specific, the constraints for any value of $r$ are given by those for $r-1$ complemented by some additional new ones. It is exactly the parameter $r$ rather than $N$ and $d$ that determines the complexity of our scientific problem. 
Since most of the essential low-energy physics is captured by the lowest two or three eigenstates, it suffices for most applications (see subsequent point (ii)) to restrict to $r \leq 3$. This allows us, in each symmetry sector, to compute the ground state and the first two excited states, and hence the two lowest excitation gaps. 
\item Since our work introduces and solves a new variant of the single-body quantum marginal problem, it contributes to the general development in the quantum information sciences \cite{KL04, DH05, KL06, CM06, Ruskai_2007, AK08, KL09, JJR12-prl, JJR12, SWK13, S15, MT17, CS17stability, WHG17, MSGLS20, SBLMS20, SP21, LCLS21, Aloy_2021, YSWCG21, HLA22, Fraser22, CLLPPS23, CL23}
that is all about describing the compatibility of reduced density matrices (quantum marginals), with ample applications in quantum information processing.

The ultimate motivation of our work, however, was to support the recent development of reduced density matrix methods, particularly ensemble density functional theory (EDFT) \cite{Theophilou79, GOK88c, TG95, YPBU17, Fromager2020-DD, Loos2020-EDFA, Cernatic21, GK21, Yang21, GKGGP23, GL23, SKCPJB24, CPSF24} and ensemble one-particle reduced density matrix functional theory ($\wb$-RDMFT) \cite{SP21, LCLS21, LS23-njp, LS23-sp}, for targeting excited states. As we will explain in detail in Sec.~\ref{sec:appl}, the solution of our (relaxed) spin-symmetry adapted $\wb$-ensemble $N$-representability problem will reveal a compact characterization of the still unknown domain of the universal interaction functionals. In that sense, our work will provide a crucial missing cornerstone for these methods.

From a general point of view, the reader may also wonder why we address the spin-symmetry adapted $N$-representability problem not on the level of the full 1RDM $\gamma$. As explained for pure states in Refs.~\cite{AK08, LMS23, LMS24-jctc}, the reason for this is that the corresponding set of 1RDMs is not invariant under unitary transformations on $\mathcal{H}_1$ and thus impossible to compute for practically relevant system sizes $N,d$. This follows from the fact that the total spin operator $\bd S^2$ does not commute with arbitrary one-particle operators.
\item The relaxed $\wb$-ensemble $N$-representability problem without spin has already been solved by some of us in Refs.~\cite{SP21, LCLS21,CLLPPS23}. Yet, this solution is not directly applicable in $\wb$-RDMFT or EDFT for the calculation of excited states. The reason for this is that common functional approximations refer to a specific spin-symmetry sector. Changing this paradigm is problematic since finding an accurate ansatz for the underlying $N$-electron ensemble state is almost impossible if one has to treat different spin states on the same footing. Accordingly, it is a formidable challenge to address the case with spin.

    Unlike the spin-independent case, the present work requires the additional integration of representation theory into the toolkit developed for solving the one-body $\bd w$-ensemble $N$-representability problem in Refs.~\cite{SP21, LCLS21, CLLPPS23}. Moreover, incorporating spin quantum numbers $S$ and $M$ necessitates a modified derivation of the vertex representation of the spectral polytope.
    A crucial difference to the spin-independent setting is that two orthonormal configuration states (see below) can now correspond to the same spectrum of the orbital 1RDM.
\end{enumerate}

\section{Solution of relaxed spin-adapted one-body $N$-representability problem\label{sec:QMP}}

The spectral set $\Sigma_{N,S}(\bd w) $ is a convex polytope as a direct consequence of the convexity theorems about moment polytopes in Refs.~\cite{Kostant73, Atiyah82, GS82}. According to the Minkowski-Weyl theorem, this implies that $\Sigma_{N,S}(\bd w) $ is equivalently characterized by its vertex representation or a minimal hyperplane representation \cite{Ziegler1994}. In this section we derive first the vertex representation and then turn it into a minimal hyperplane representation by using tools from discrete geometry. The general scheme uses similar mathematical tools as presented in detail for the spin-independent case in Refs.~\cite{LCLS21, CLLPPS23, CL23}. Nevertheless, there are several crucial differences compared to the spin-independent setting. This includes in particular a different construction of the Hasse diagram of partially ordered configurations and its dependence on the quantum numbers $S,M$. Therefore, we focus in the following on these differences in the solution to the $\bd w$-ensemble $(N,S,M)$-representability problem arising from fixing the spin quantum numbers $S,M$. The derivation of the spin adapted $\bd w$-ensemble $(N,S,M)$-representability constraints of the orbital 1RDM provides insights into the geometric structure of the respective convex polytopes $\Sigma_{N,S}(\bd w) $. Readers only interested in the constraints themselves might directly jump to Sec.~\ref{sec:examples}.

\subsection{Hasse diagram of partially ordered configurations\label{sec:Hasse}}

As a consequence of the duality principle for compact convex sets and the Minkowski-Weyl theorem, the spectral set $\Sigma_{S, M}(\bd w)$ is fully characterized by the intersection of its supporting hyperplanes \cite{rockafellar2015, Ziegler1994}.
A supporting hyperplane of the polytope $\Sigma_{N,S}(\bd w)$ at a point $\bd\lambda^\prime\in \Sigma_{N,S}(\bd w)$ is a hyperplane that satisfies $a^\mathrm{T} \cdot\bd\lambda \leq a^\mathrm{T}\cdot\bd\lambda^\prime$ for all $\bd\lambda\in \Sigma_{N,S}$, where $\bd a$ denotes the normal vector of the hyperplane \cite{Ziegler1994}.
Moreover, the dual variable of the orbital 1RDM $\gamma_l$ on $\mathcal{H}_1^{(l)}$ is the orbital one-particle Hamiltonian
\begin{equation}\label{eq:hl}
h_l \equiv \sum_{i,j=1}^d h_{ij}E_{ij}
\end{equation}
and, in analogy to Refs.~\cite{SP21, LCLS21}, we can assume without loss of generality that $h_l$ and $\gamma_l$ share a common eigenbasis. We denote the vector of eigenvalues of $h_l$ by $\bd h\in \RR^d$.
In combination with the aforementioned duality principle, this means that we have to minimize linear functionals $\langle \bd h, \bd\lambda\rangle=\mathrm{const.}$ over the set $\Sigma_{N,S}$ for all possible normal vectors $\bd h$ of the corresponding hyperplanes.
Due to the unitary invariance of the set $\ebw$ we can further restrict to those orbital one-particle Hamiltonians $h_l$ with a increasingly ordered spectrum $h_1^{(l)}\leq h_2^{(l)}\leq \ldots\leq h_d^{(l)}$ and discuss only $\Sigma^\downarrow_{N,S}(\bd w)$ instead of $\Sigma_{N,S}(\bd w)$, as introduced in Sec.~\ref{sec:science}.

The supporting hyperplanes are calculated by considering a so-called Hasse diagram of partially ordered configurations. To this end, we first introduce the set $\mathcal{I}_{N,S,M}$ of configuration states $\ket{\bd i}$ that constitute a basis for the $N$-particle Hilbert $\mathcal{H}_N^{(S,M)}$ of states with fixed quantum numbers $S,M$. The set $\mathcal{I}_{N,S,M}$ can be constructed from chosen orthonormal bases $\mathcal{B}_l, \mathcal{B}_s$ of $\mathcal{H}_1^{(l)}, \mathcal{H}_1^{(s)}$ by constructing a Verma basis of the irreducible representation space as explained in Refs.~\cite{LMNP86, PP18} (see also the textbook \cite{Hall15}). The spatial (or, also called orbital) configurations $\bd i = (i_1, i_2, \ldots, i_N)$ denote the spatial orbitals $\ket{i}, i=1, \ldots, d$ occupied by the $N$ fermions. Due to the spin degree of freedom each $i$ can appear at most twice in a configuration $\bd i$. The energetically lowest configuration $\ket{\bd i_0}\in \mathcal{I}^N_{S,M}$ is the highest weight state $\ket{\Lambda}$ \cite{Hall15, LMS23}, which is unique according to Cartan's theorem \cite{Hall15}. By definition, the highest weight state $\ket{\Lambda}$ is annihilated by all positive root operators $E_{ij}, j<i$ with $E_{ij}$ defined in Eq.~\eqref{eq:Eij}. For $S$ and the special case $M=S$ as good quantum numbers, a state $\ket{\Lambda_{S, M=S}}$ is the unique highest weight state if and only if
\begin{equation}
\forall j<i:\,\,E_{ij}\ket{\Lambda_{S, M=S}} = 0\quad \wedge\quad S^{+}\ket{\Lambda_{S, M=S}} = 0 \,,
\end{equation}
where $S^+= \sum_{i=1}^dS_i^+$.
Therefore, the highest weight state for $\mathcal{H}_N^{(S, S)}$ is given by
\begin{equation}\label{eq:Lambda-hwv}
\ket{\Lambda_{S, M=S}} = \ket{\underbrace{1\alpha, 1\beta, \ldots, K\alpha, K\beta}_{N-2S}, \underbrace{(K+1)\alpha, \ldots, J\alpha}_{2S}, 0, \ldots}\,,
\end{equation}
where we introduced
\begin{equation}\label{eq:kj}
K=\frac{N-2S}{2}\,,\quad J=\frac{N+2S}{2}\,.
\end{equation}
The highest weight state $\ket{\Lambda_{S,M}}$ of $\mathcal{H}_N^{(S,M)}$ follows then from $\ket{\Lambda_{S, M=S}}$ by a finite number of applications of the negative root operator $S^-=\sum_{i=1}^dS_i^-$ until the desired $M$ is reached. In particular, in contrast to Slater determinants, as used in Refs.~\cite{SP21, LCLS21}, the weights of the configurations $\bd i$ for $(S,M)$ do not necessarily correspond to vertices of the set $\Sigma_{N,S}(\bd w = (1, 0, \ldots))$. Here, the word weight originates from a representation theoretical point of view and refers to the natural occupation number vector of the orbital 1RDM of a configuration state.
If these weights are not extremal elements of the set $\Sigma_{N,S}(\bd w)$, this implies that they do not correspond to non-degenerate ground states of a non-interacting Hamiltonian \eqref{eq:hl}. Still they have to be taken into account when constructing the excitation pattern. This will become important in Sec.~\ref{sec:v-repr}.
Beyond the ground state configuration, the partial ordering of the configurations $\bd i$ is defined equivalently to Ref.~\cite{LCLS21}: $\bd i\leq \bd j$ if and only if $i_{k_1} + \ldots+  i_{k_N}\leq j_{k_1}+ \ldots+\leq j_{k_N}$.

Before we work out the Hasse diagram of partially ordered configurations, we observe that the operators $E_{ij}$ are singlet irreducible tensor operators of the group $SU(2)$. Therefore, the Wigner-Eckart theorem \cite{Hall15} implies that the expectation value of the 1RDM matrix elements $(\gamma_l)_{ij}$ for a given state $\ket{\Psi_{S,M}}\in \mathcal{H}_N^{(S,M)}$ equal the reduced matrix elements (see also Eq.~\eqref{eq:1RDM}). This implies that despite the configuration states for $S,M$ differ for different $M$, their configurations $\bd i$ are equal and we can restrict without loss of generality to $S=M$ for the construction of the Hasse diagram of partially ordered orbital configurations. On the level of the spectral set this implies that $\Sigma_{N,S}(\bd w)=\Sigma_{S,M^\prime}(\bd w)$ for all $M, M^\prime=-S, \ldots, S$.
Furthermore, we observe that the structure of Hasse diagram is effectively independent of $S, N, d$ if they satisfy the three conditions
\begin{align}
d- \frac{N+2S}{2}&\geq r-1\,,\label{eq:stability1}\\
\quad \frac{N-2S}{2}&\geq r-1\,,\label{eq:stability2}\\
\quad 2S&\geq r-1\,,\label{eq:stability3}
\end{align}
where $r$ denotes the number of non-vanishing entries in $\wb$. These three stability conditions are important as the resulting Hasse diagram and, therefore, the solution to the $N$-representability problem is effectively independent of $S,N, d$ as long as they are fulfilled. In particular, this allows us to derive the relaxed $\bd w$-ensemble $(N,S,M)$-representability constraints for generic $N,S,d$ in Sec.~\ref{sec:H-repr}.

\begin{figure}[tb]
\centering
\frame{\includegraphics[width=\linewidth]{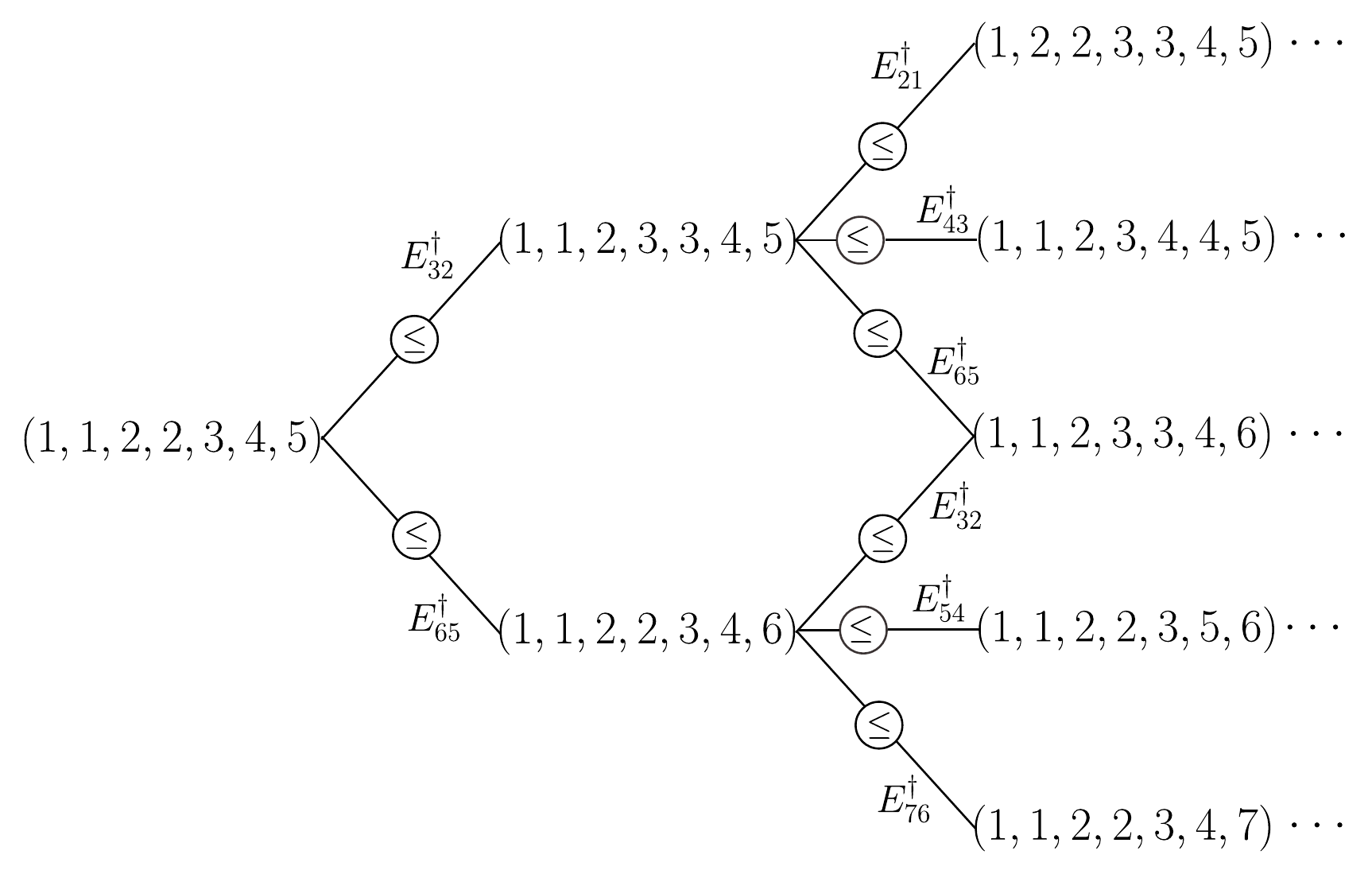}}
\caption{Illustration of the Hasse diagram of partially ordered configurations for $r\leq 3$ satisfying the stability conditions for $S,N,d$ with $N=7, S=3/2$. (see text for more explanations). \label{fig:spectrum-generic}}
\end{figure}
\begin{figure*}[tb]
\centering
\frame{\includegraphics[width=0.9\linewidth]{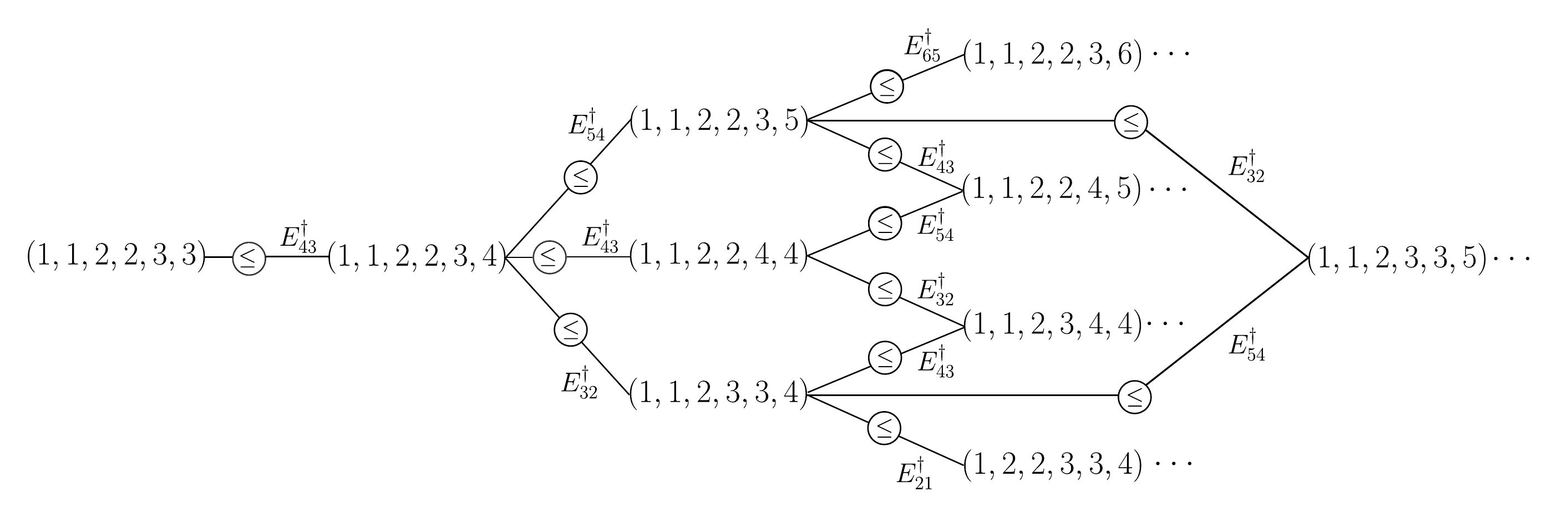}}
\caption{Illustration of Hasse diagram of partially ordered configurations for a singlet and $N=6$. \label{fig:singlet-spectrum}}
\end{figure*}
To build the Hasse diagram, we start from the ground state configuration state, namely the highest weight state $\ket{\Lambda_{S,S}}$. We then determine all single negative root operators $E_{ij}^\dagger, j<i$ that satisfy $E_{ij}^\dagger\ket{\Lambda_{S,M}}\neq 0$ with $i-j=1$. The construction of the next higher orbital configurations follows analogously.
The restriction on $i-j=1$ guarantees that we find the configuration corresponding to the first excited state, i.e., that the partial ordering condition is satisfied. For $N,S,d$ satisfying the stability conditions Eqs.~\eqref{eq:stability1}-\eqref{eq:stability3} this implies that there are two possible configurations of first excitation. In fact, this is in contrast to the spin-independent case where the first excitation is always unique as explained in Refs.~\cite{SP21, LCLS21,CLLPPS23}. We use the resulting Hasse diagram in the next section to construct the vertex representation of the spectral polytope $\Sigma_{N,S}(\bd w)$.

We illustrate the Hasse diagram of partially ordered configurations for generic $S,d,N$ satisfying Eqs.~\eqref{eq:stability1}-\eqref{eq:stability3} for $N=7, S=3/2$ in Fig.~\ref{fig:spectrum-generic}. The orbital configuration of the unique highest weight state $\ket{\Lambda_{S,S}} = \ket{1\alpha, 1\beta, 2\alpha, 2\beta, 3\alpha,4\alpha, 5\alpha} $ is given by $\bd i_0 = (1,1,2,2,3,4,5)$. The two orbital configurations $\bd i=(1,1,2,3,3,4,5), (1,1,2,2,3,4,6)$ of the first excitation are obtained by acting with the respective negative root operators $E_{32}^\dagger, E_{65}^\dagger$ on the highest weight state $\ket{\Lambda_{S,S}}$.
Moreover, the singlet case, $S=0$, does not satisfy the third condition \eqref{eq:stability3} for $r\geq 2$. Therefore, we still have a unique first excited state for $S=0$ and the Hasse diagram for a singlet with $N=6$ illustrated in Fig.~\ref{fig:singlet-spectrum} indeed differs from Fig.~\ref{fig:spectrum-generic}.
Due to this difference between $S=0$ and generic $S$ satisfying the stability conditions Eqs.~\eqref{eq:stability1}-\eqref{eq:stability3}, we also have to derive the corresponding relaxed orbital one-body $\bd w$-ensemble $N$-representability constraints for the two cases separately in Sec.~\ref{sec:examples}. The two Hasse diagrams in Figs.~\ref{fig:spectrum-generic} and \ref{fig:singlet-spectrum} provide the key ingredient for the derivation of these constraints as explained in the following.

\subsection{Vertex representation\label{sec:v-repr}}

Equipped with the Hasse diagram determined in the previous section, we are now in a position to derive the vertex representation of $\Sigma_{N,S}(\bd w)$.
The new stability conditions in Eqs.~\eqref{eq:stability1}-\eqref{eq:stability3} and the different construction of the Hasse diagram of the orbital configurations through the positive root operators $E_{ij}$ of $\mathfrak{u}(d)$ are the first two important differences to the spin-independent case discussed in Refs.~\cite{SP21, LCLS21, CLLPPS23}.
In the following, we will discuss the third main difference, namely that the configurations $\bd i$ can have multiplicities larger than one, which is not possible for Slater determinants as in the spin-independent setting of Refs.~\cite{SP21, LCLS21, CLLPPS23}. The multiplicity $m_{\bd i}$ of a configuration $\bd i$ is given by (e.g., see the textbook \cite{Pauncz12})
\begin{equation}
m_{\bd i} = \begin{pmatrix}
N_u\\\frac{N_u}{2}-S
\end{pmatrix}
- \begin{pmatrix}
N_u\\\frac{N_u}{2}-S-1
\end{pmatrix}\,,
\end{equation}
where $N_u$ denotes the number of unpaired electrons in $\bd i$.
Alternatively, the multiplicity of a configuration can be calculated using Kostant's multiplicity formula  \cite{K59, Hall15}. Moreover, the configuration of the highest weight state always has multiplicity one due to the uniqueness theorem of the highest weight proved by Cartan \cite{Hall15}. The other configurations in Figs.~\ref{fig:spectrum-generic} and \ref{fig:singlet-spectrum} also have multiplicity one. However, configurations with $m_{\bd i}>1$ occur when including the configurations of higher excited states in those Hasse diagram. For instance, in the case of $N=9, S=3/2$ in Fig.~\ref{fig:singlet-spectrum}, the configuration $\bd i = (1,1,2,2,3,4,5,6,7)$ with $m_{\bd i} = 4$ will occur if more orbital configurations are included in the Hasse diagram. At the same time, it can be easily verified that the three stability conditions ensure that $m_{\bd i}=1$ for $S,N,d$ satisfying Eqs.~\eqref{eq:stability1}-\eqref{eq:stability3} for a given fixed rank $r$ of the $N$-fermion states $\Gamma$.

The multiplicities of the configurations are important as they have to be taken into account when constructing the so-called lineups $l$ as explained in the following. A lineup $l$ of length $r$ is a sequence of the $r$ largest configurations that respects the partial order between them \cite{LCLS21, CLLPPS23, CL23}. The possible lineups are determined by the Hasse diagram as follows: consider first $r=2$, i.e., rank-2 states with fixed spectrum $\bd w$ on the $N$-particle level.
As explained in Sec.~\ref{sec:H-repr}, there are two possible orbital configurations for the second lowest orbital configuration for $r=2$. This leads to $R=2$ lineups for $r=2$.
For the examples of Hasse diagrams in Figs.~\ref{fig:spectrum-generic} and \ref{fig:singlet-spectrum} this leads to two lineups $l_1, l_2$ for Fig.~\ref{fig:spectrum-generic}, while we have only one lineup $l_1$ for Fig.~\ref{fig:singlet-spectrum}. The number of distinct lineups for fixed $r$ is denoted by $R(r)$.
As the multiplicity $m_{\bd i}$ of a configuration corresponds to the number of orthogonal configuration states mapping to it, configurations have to be taken into account $m_{\bd i}$-many times.

\begin{figure}[tb]
\frame{\includegraphics[width=\linewidth]{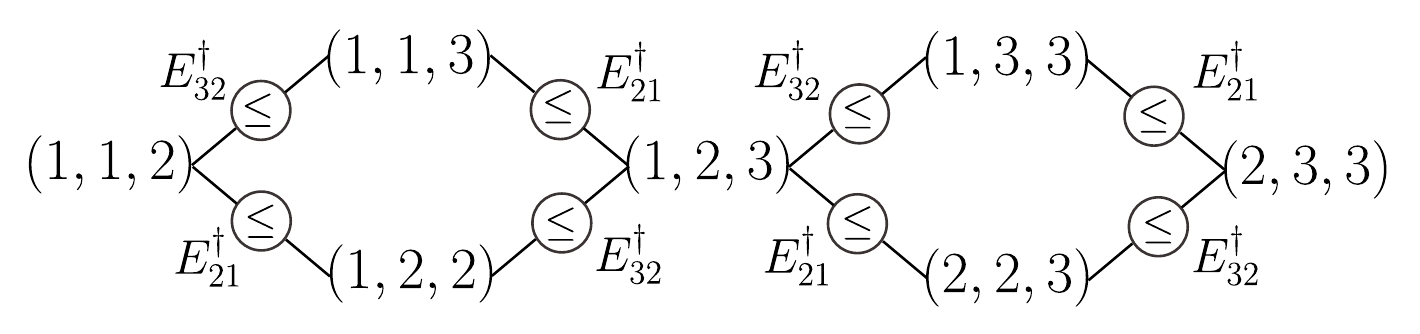}}
\caption{Hasse diagram of partially ordered configurations for $N=d=3$ and $S, M=1/2$.\label{fig:BD}}
\end{figure}

As a simple example, we consider the setting of $N=3$ fermions in $d=3$ orbitals (recall that $d=\mathrm{dim}(\mathcal{H}_1^{(l)})$) and $S=M=1/2$. The respective Hasse diagram is shown in Fig.~\ref{fig:BD}. The dimension of $\mathcal{H}_N^{(S,M)}$ is $D_{S,M}=8$ but there are only seven configurations as the configuration $\bd i =(1,2,3)$ has multiplicity two. Thus, for $r\geq 5$, the configuration $\bd i =(1,2,3)$ has to be taken into account twice before considering $(1,3,3),(2,2,3)$. This example leads for $r=5$ to the two lineups
\begin{align}
&l_1\!:(1,1,2)\to(1,1,3)\to (1,2,2)\to(1,2,3)\to(1,2,3)\,,\nonumber\\
&l_2\!:\,(1,1,2)\to(1,2,2)\to (1,1,3)\to(1,2,3)\to(1,2,3)\,.
\end{align}
Moreover, the vector of eigenvalues $\bd \lambda$ of $\gamma_l$ that corresponds to $\bd i = (1,2,3)$ is not extremal in the set $\Sigma_{N,S}(\bd w_0)$, $\bd w_0=(1,0,\ldots)$ \cite{LMS23}. This leads to a further important difference to the spin-independent case, namely that going from $r=2$ to $r=3$ in this example does not yield additional constraints beyond those of $r=2$.

The remaining part of the derivation of the vertex representation is analogous to the spin-independent setting and we refer the reader for more details to Refs.~\cite{SP21, LCLS21}. Each lineup $l_i, i=1, \ldots,R(r)$ of length $r$ then leads to a generating vertex $\bd v^{(i)}$ according to
\begin{equation}\label{eq:vertex}
\bd v^{(i)} = \sum_{J=1}^rw_J\bd n^{(J)}\,,
\end{equation}
where $\bd n^{(J)} = \mathrm{spec}\left(\mu(\ket{\bd j_J}\!\bra{\bd j_J})\right)$ with $n^{(J)}_j \in \{0, 1, 2\}$.
The vertices $\bd v^{(i)}, i=1, \ldots, R(r)$ are called generating vertices as all other vertices of $\Sigma_{N,S}(\bd w)$ follow from permutations of their entries. In particular, the vertex representation of $\Sigma_{N,S}(\bd w)$ is given by
\begin{equation}\label{eq:Sigma-vrep}
\Sigma_{N,S}(\bd w) = \mathrm{conv}\left(\left\{\pi(\bd v^{(i)})\,|\,i=1, \ldots, R(r), \pi\in \mathcal{S}_d\right\}\right)\,.
\end{equation}

\subsection{Minimal hyperplane representation\label{sec:H-repr}}

In this section, we translate the vertex representation of $\Sigma_{N,S}(\wb)$ to a minimal hyperplane representation. We use the same techniques that were worked out in detail in Ref.~\cite{CLLPPS23}. To keep the work self-contained, we therefore only briefly recap the most important concepts from discrete geometry that are required in Sec.~\ref{sec:examples} to derive the relaxed orbital one-body $\bd w$-ensemble $(N,S,M)$-representability constraints.
For a single generating vertex this is trivial since the linear constraints follow directly from a theorem by Rado \cite{R52}. However, the number of lineups and, thus, the number of generating vertices grows with $r$ which makes the task highly non-trivial.

Let us assume that $N, S, d$ satisfy the stability conditions in Eqs.~\eqref{eq:stability1}-\eqref{eq:stability3} for a given $r$.
The first step to arrive at the hyperplane representation of $\Sigma_{N,S}(\bd w)$ is to determine all lineups $l_i, i=1, \ldots, R(r)$ and calculate the corresponding $R(r)$ generating vertices as described in Sec.~\ref{sec:v-repr}. A linear inequality is called valid if it is satisfied by all points in the polytope $\Sigma_{N,S}(\bd w)$.
The normal fan $\mathcal{N}(\Sigma_{N,S}(\bd w))$ of $\Sigma_{N,S}(\bd w)$ is determined from the normal cones of the vertices of $\Sigma_{N,S}(\bd w)$ derived in Sec.~\ref{sec:v-repr}.
The facets of $\Sigma_{N,S}(\bd w)$ provide all missing normal cones of $\Sigma_{N,S}(\bd w)$ and yield the rays used to calculate the facet-defining inequalities. Due to permutation invariance of $\Sigma_{S, M}(\bd w)$ it is further sufficient to study the fundamental fan $\mathcal{N}_f(\Sigma_{S, M}(\bd w))$, which is given by the intersection of the normal fan with the fundamental chamber $\Phi_d$, i.e. $\mathcal{N}_f(\Sigma_{N, S}(\bd w)) =  \mathcal{N}(\Sigma_{N, S}(\bd w))\cap \Phi_d$. The fundamental chamber $\Phi_d$ is defined as the set
\begin{equation}
\Phi_d \equiv \left\{\bd\eta\in (\RR^d)^*\,|\,\bd\eta = \bd\eta^\downarrow\right\}
\end{equation}
of linear functionals on the dual space $ (\RR^d)^*$.
Furthermore, we introduce the dual fundamental basis $\mathcal{B}_f = \{\bd f_i\}_{i=1}^d$ whose basis vectors are given by
\begin{equation}\label{eq:f-dual}
\bd f_i\equiv\sum_{j=1}^i\bd\varepsilon_j\,,\quad \forall\,i=1, \ldots, d\,,
\end{equation}
where $\bd\varepsilon_i$ are the elements of the dual elementary basis. The set of linear constraints is expressed as $A\cdot\bd\lambda^\downarrow\leq B\cdot\bd w$, where $A,B$ are coefficient matrices and each row determines one linear constraint that has to be satisfied by any $\bd\lambda\in\Sigma_{N,S}(\bd w) $.
For each lineup $l_k$ we then determine the set of fundamental linear functionals $\bd \eta$ uniquely maximised at the corresponding generating vertex $\bd v^{(k)}$ and determine from that the rays of $\mathcal{N}_f(\Sigma_{N, S}(\bd w))$. The resulting rays determine the left hand-side of the linear constraints in our hyperplane representation of $\Sigma_{N, S}(\bd w)$, i.e., the matrix $A$, and evaluating them on the vertices $\bd v^{(k)}, k=1, \ldots, R$ yields the right hand-side of the inequalities.
Due to the intersection with the fundamental chamber this procedure might lead to redundant inequalities which are removed afterwards. Thus, the last step is to determine the minimal hyperplane representation of $\Sigma_{N, S}(\bd w)$ by identifying the redundant inequalities. We illustrate these concepts and derive the resulting inequalities for different $r$ and $S$ and generic $N,d$ in Sec.~\ref{sec:examples}. This will eventually lead to the sought-after spin-dependent $\bd w$-ensemble orbital one-body $(N,S,M)$-representability constraints.

\begin{figure*}[htb]
\includegraphics[width=\linewidth]{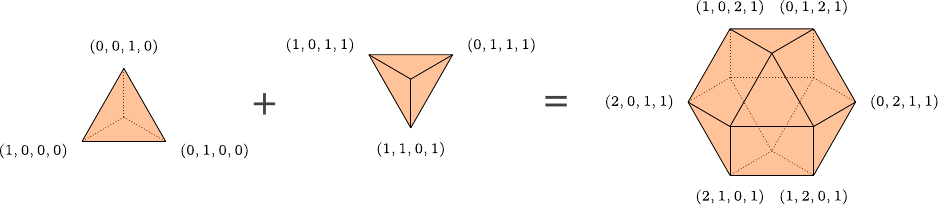}
\caption{The spectral polytope $\Sigma_{N,S}(\bd w_0)$ with $\bd w_0=(1, 0, \ldots)$ for $S=1, d=N=4$ follows as the Minkowski sum of the permutohedra of $\bd v_1=(1, 0, 0,0)$ and $\bd v_2=(1,1,1,0)$ of $N=1$ and $N=3$ fermions, respectively. The value of $\lambda_4=N-\sum_{i=1}^3\lambda_i$ is fixed through the normalization of $\gamma_l$ to the total particle number $N$.  \label{fig:minkowski}}
\end{figure*}

\section{Spin-adapted exclusion principle constraints for mixed states\label{sec:examples}}

In this section we explicitly derive the relaxed $\bd w$-ensemble one-body $N$-representability constraints for different values of the total spin quantum number $S$.

\subsection{Geometrical aspects of spin-adapted Pauli constraints\label{sec:r1}}

Before deriving the spin-adapted $\bd w$-ensemble one-body $N$-representability constraints for $r\geq 2$, we discuss in this section the case $r=1$ from a geometric perspective. The unique generating vertex of $\Sigma_{N, S}(\bd w_0), \bd w_0=(1, 0, \ldots)$ follows as the highest weight as explained in Sec.~\ref{sec:Hasse} as (recall Eqs.~\eqref{eq:Lambda-hwv},\eqref{eq:kj})
\begin{equation}\label{eq:v-hw}
\bd v = (\underbrace{2, \ldots, 2}_{(N-2S)/2}, \underbrace{1, \ldots, 1}_{2S}, \underbrace{0, \ldots, 0}_{d-(N+2S)/2})
\end{equation}
which leads to the so-called spin-adapted Pauli constraints \cite{LMS23, LMS24-jctc}
\begin{align}\label{eq:spin-Pauli}
\lambda_1^\downarrow &\leq 2\,,\nonumber\\
&\vdots\nonumber\\
\sum_{i=1}^{K}\lambda_i^\downarrow &\leq N-2S\,,\nonumber\\
\sum_{i=1}^{K+1}\lambda_i^\downarrow &\leq N-2S+1\,,\nonumber\\
&\vdots\nonumber\\
\sum_{i=1}^{J}\lambda_i^\downarrow &\leq N\,.
\end{align}

We are interested in the geometric structure of the permutohedron $\Sigma_{N, S}(\bd w_0)$. To this end, we define the two vertices
\begin{align}
\bd p^{(1)} &= (\underbrace{1, \ldots, 1}_{(N-2S)/2}, \underbrace{0, \ldots, 0}_{d-(N-2S)/2})\nonumber\\
\bd p^{(2)} &= (\underbrace{1, \ldots, 1}_{(N+2S)/2}, \underbrace{0, \ldots, 0}_{d-(N+2S)/2})\,.
\end{align}
and the corresponding permutohedra
\begin{equation}
P^{(i)}_{N,S} = \mathrm{conv}\left(\left\{\pi(\bd p^{(i)}) \,|\,\pi\in \mathcal{S}_d  \right\}\right)\,.
\end{equation}
Thus, $P^{(1)}_{N,S} $ is nothing else than the Pauli simplex for the setting of $N_\beta \equiv K=(N-2S)/2$ spin-$\beta$ electrons, while $P^{(2)}_{N,S}$ corresponds to the Pauli simplex for $N_\alpha \equiv J = (N+2S)/2$ spin-$\alpha$ electrons.
Then, $\Sigma_{N, S}(\bd w_0)$ follows as the Minkowski sum of the two permutohedra $P_{N,S}^{(1)}$, $P_{N,S}^{(2)}$ according to
\begin{align}\label{eq:minkoswki-Pauli}
\Sigma_{N, S}(\bd w_0) &= P_{N,S}^{(1)} + P_{N,S}^{(2)}\\
&= \left\{ \bd x_1+\bd x_2\,|\,\bd x_1\in P_{N,S}^{(1)}, \bd x_2\in P_{N,S}^{(2)}\right\} \,.\nonumber
\end{align}
We illustrate this Minkowski sum for $N=d=4, S=1$ in Fig.~\ref{fig:minkowski}, where $\lambda_4 = N-\sum_{i=1}^3\lambda_i$ is fixed through the normalization condition. In that case, $\bd p^{(1)} = (1, 0, 0,0)$ and $\bd p^{(2)} = (1,1,1,0)$. Therefore, $\Sigma_{4, 1}(\bd w_0)$ with generating vertex $\bd v = (2,1,1,0)$ (right) is nothing else than the Minkowski sum of $P^{(1)}_{4,1}$ (left) and $P^{(2)}_{4,1}$ (middle).

As $\Sigma_{N, S}(\bd w_0)$ is the sum of two hypersimplices $P^{(1)}_{N,S}, P^{(2)}_{N,S}$, it has a richer geometric structure than the Pauli hypersimplex (for spinless fermions), which is the permutohedron of the Hartree-Fock point $\bd v^\mathrm{HF} = (1, \ldots, 1, 0, \ldots)$. In fact, the volume of a hypersimplex is well-known since more than $100$ years \cite{Laplace}, while the volume of the sum of hypersimplices is in general more complicated to determine \cite{Postnikov09, Liu16}.

\subsection{Constraints for generic $S, N, d$ and $r=2$\label{sec:r2}}

In the following, we derive the hyperplane representation of the spectral polytope $\Sigma_{N,S}(\bd w)$ for $r=2$. We consider generic $S,N,d$ satisfying the three stability constraints in Eqs.~\eqref{eq:stability1}-\eqref{eq:stability3}. Thus, we consider the setting illustrated in Fig.~\ref{fig:spectrum-generic}, where there are two possible first excited states and, thus, two corresponding orbital configurations. These two orbital configurations are obtained from the highest weight state $\ket{\Lambda_{S,M=S}}$ in Eq.~\eqref{eq:Lambda-hwv} by applying the two negative root operators $E_{K+1,K}^\dagger, E_{J+1,J}^\dagger$, respectively. For the example of $N=7, S=3/2$ in Fig.~\ref{fig:spectrum-generic} these are the operators $E_{32}^\dagger, E_{65}^\dagger$ with $K, J$ defined in Eq.~\eqref{eq:kj}.

These two configuration are then used as described in Sec.~\ref{sec:v-repr} to derive the vertex representation of the spectral polytope $\Sigma_{N,S}(\bd w)$ in Eq.~\eqref{eq:Sigma-vrep} for $r=2$. This leads to the following two generating vertices (see Appendix \ref{app:genS} for a detailed derivation)
\begin{align}\label{eq:v-r2}
\bd v^{(1)} &= (\underbrace{2, \ldots, 2}_{\frac{N-2S}{2}}, \underbrace{1, \ldots, 1}_{2S-1}, w_1, 1-w_1, 0, \ldots)\,,\nonumber\\
\bd v^{(2)} &= (\underbrace{2, \ldots, 2}_{\frac{N-2S}{2}-1}, 1+w_1, 2-w_1,\underbrace{1, \ldots, 1}_{2S-1}, 0, \ldots)\,.
\end{align}
To translate this vertex representation of $\Sigma_{N,S}(\bd w)$ in Eq.~\eqref{eq:Sigma-vrep} into a hyperplane representation, we follow in Appendix \ref{app:genS} the formalism outlined in Sec.~\ref{sec:H-repr}. The vertices in Eq.~\eqref{eq:v-r2} are used in this derivation to calculate from the fundamental rays of the normal cone the right hand-side of the respective inequalities. This eventually yields the spin-adapted Pauli constraints in Eq.~\eqref{eq:spin-Pauli} for $r=1$ and the new and non-trivial constraint for $r=2$
\begin{align}\label{eq:r2-w-constraints}
2\sum_{i=1}^K \lambda_i^\downarrow+\sum_{j=K+1}^J\lambda_j^\downarrow &\leq 2(N-S)-1+w_1\,,
\end{align}
which indeed depends on $\wb$ as well as on the quantum numbers $S,N$. Its independence from $M$ is a consequence of the Wigner-Eckart theorem as explained in Sec.~\ref{sec:Hasse}. The additional $\bd w$-dependent constraint for $r=2$ constitutes the first non-trivial exclusion principle constraint for $\bd w$-ensembles with fixed good quantum numbers $S,M$. For $r=3$ discussed in the next section, we again have the constraints from $r=1,2$ in Eq.~\eqref{eq:r2-w-constraints} together with a finite number of additional constraints.

\begin{figure*}[tb]
\includegraphics[width=\linewidth]{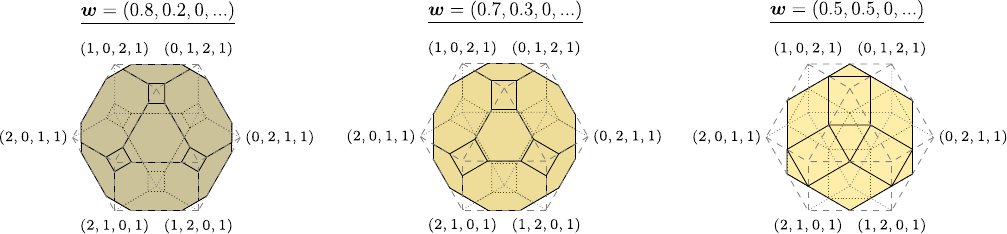}
\caption{Illustration of the spectral polytope $\Sigma_{N,S}(\bd w)$ for $N=d=4, S=1$ and three different vectors $\bd w$ with $r=2$ non-vanishing weights. The value of $\lambda_4 = N-\sum_{i=1}^3\lambda_i$ is fixed through $\lambda_1, \lambda_2, \lambda_3$.
The spectral set $\Sigma_{N,S}(1,0,\ldots)$ that corresponds to the Pauli constraints \eqref{eq:spin-Pauli} is shown in gray. The additional $\bd w$-dependent constraint for $r=2$ removes the corners of the spin-adapted Pauli spectral polytope and its $\bd w$-dependence reduces the size of $\Sigma_{N,S}(\bd w)$ as the mixedness of $\bd w$ increases. \label{fig:poly_S1}}
\end{figure*}

We illustrate the more restrictive spin-adapted $\bd w$-ensemble $(N,S)$-representability constraints for three different choices of the weight vector $\bd w$ and $N=d=4, S=1$ in Fig.~\ref{fig:poly_S1}. For these values of $N, d, S$ the stability conditions in Eqs.~\eqref{eq:stability1}-\eqref{eq:stability3} are satisfied.
Due to the normalization of the orbital 1RDM $\gamma_l$ to the total particle number $N$, the forth dimension can be discarded as $\lambda_4 = 4-\sum_{i=1}^3\lambda_3$.
The spectral polytope arising from the spin-adapted Pauli constraints in Refs.~\cite{LMS23, LMS24-jctc} is shown by the gray dashed lines and contains $\Sigma_{N,1}(\bd w)$. Moreover, we observe an inclusion relation with respect to $\bd w$ in analogy to Refs.~\cite{LCLS21, CLLPPS23}, namely $\Sigma(\bd w^\prime)\subseteq \Sigma(\bd w)$ if and only if $\bd w^\prime\prec\bd w$. For example, this implies that the right polytope for $\bd w^\prime = (0.5,0.5,0, \ldots)$ is contained in the left polytope with $\bd w=(0.7, 0.3, 0, \ldots)$.

\subsection{Constraints for generic $S, N, d$ and $r=3$\label{sec:r3}}

In the following, we derive the constraints for $r=3$ and generic $S,N,d$ satisfying the stability conditions in Eqs.~\eqref{eq:stability1}-\eqref{eq:stability3}. We will see that this leads again to the spin-adapted Pauli constraints \eqref{eq:spin-Pauli}, the new spin-adapted $\bd w$-constraint in Eq.~\eqref{eq:r2-w-constraints} and additional new constraints.
Moreover, the hyperplane representation for $r=3$ has to reduce to the one for $r=2$ in the limit $w_3\rightarrow 0$.

For $r=3$, the vertex representation of $\Sigma_{N,S}(\bd w)$ for generic $S, N, d$ follows from the respective Hasse diagram of partially ordered configurations as (see Appendix \ref{app:genS})
\begin{widetext}
\begin{align}\label{eq:v-r3}
\bd v^{(1)} &= (\underbrace{2, \ldots, 2}_{\frac{N-2S}{2}-1}, 2-w_2, 1+w_2, \underbrace{1, \ldots, 1}_{2S-2}, w_1+w_2, 1-w_1-w_2, 0, \ldots) \,,\nonumber\\
\bd v^{(2)} &= (\underbrace{2, \ldots, 2}_{\frac{N-2S}{2}-1}, 1+w_1+w_2, 2-w_1-w_2, \underbrace{1, \ldots, 1}_{2S-2}, 1-w_2, w_2, 0, \ldots)\,, \nonumber\\
\bd v^{(3)} &= (\underbrace{2, \ldots, 2}_{\frac{N-2S}{2}-1}, 1+w_1, 1+w_2, 2-w_1-w_2,  \underbrace{1, \ldots, 1}_{2S-2},0, \ldots)\,,\nonumber\\
\bd v^{(4)} &= (\underbrace{2, \ldots, 2}_{\frac{N-2S}{2}-2}, 1+w_1+w_2, 2-w_2, 2-w_1, \underbrace{1, \ldots, 1}_{2S-1}, 0, \ldots) \,,\nonumber\\
\bd v^{(5)} &=(\underbrace{2, \ldots, 2}_{\frac{N-2S}{2}}, \underbrace{1, \ldots, 1}_{2S-1}, w_1, w_2, 1-w_1-w_2, 0, \ldots)\,,\nonumber\\
\bd v^{(6)} &=(\underbrace{2, \ldots, 2}_{\frac{N-2S}{2}}, \underbrace{1, \ldots, 1}_{2S-2}, w_1+w_2, 1-w_2, 1-w_1, 0, \ldots)\,.
\end{align}
\end{widetext}
Calculating the rays of the normal fan as explained in Sec.~\ref{sec:H-repr} eventually yields the constraints in Eq.~\eqref{eq:r2-w-constraints} and we derive in Appendix \ref{app:genS} the additional two constraints that are characteristic for $r=3$. Thus, for $r=3$ there are three $\bd w$-dependent relaxed orbital one-body $(N,S)$-representability constraints, namely
\begin{align}\label{eq:r3-constraints}
&2\sum_{i=1}^{K}\lambda_i^\downarrow+\sum_{i=K+1}^{J}\lambda_i^\downarrow\leq 2(N-S)-1+w_1 \,,\nonumber\\
&3\sum_{i=1}^{K-1}\lambda_i^\downarrow + 2\sum_{i=K}^{K+1}\lambda_i^\downarrow+\sum_{i=K+2}^J\lambda_i^\downarrow\nonumber\\
&\qquad\qquad \leq 3N-4S-2+w_1+w_2\,,\nonumber\\
&3\sum_{i=1}^{K}\lambda_i^\downarrow + 2\sum_{i=K+1}^{J-1}\lambda_i^\downarrow+\sum_{i=J}^{J+1}\lambda_i^\downarrow\nonumber\\
&\qquad\qquad \leq 3N-2S-2+w_1+w_2\,.
\end{align}
This illustrates again the \emph{hierarchy} in $r$ of the exclusion principle constraints: In general, the constraints for any value of $r$ are given by those for $r-1$ complemented by some additional new ones. This is also consistent with the crucial fact that the polytope for the case of $r$ non-vanishing weights simplifies to the one with $r-1$ non-vanishing weights in any limit with $w_r \rightarrow 0$. Moreover, the three stability conditions \eqref{eq:stability1}-\eqref{eq:stability3} ensure that the constraints in Eq.~\eqref{eq:r3-constraints} are effectively form-independent of $N, S, d$. In particular, this implies that they do not really dependent on the size $d$ of the underlying basis set used in quantum chemical calculations and the constraints are known even in the complete basis set limit. Moreover, this form-independence of the exclusion principle constraints of $N, S, d$ is in striking contrast to Klyachko's solution to the pure state $N$-representability problem \cite{AK08}, which strongly depends on the system size. Thereby, we eventually demonstrate that the constraints in Eq.~\eqref{eq:r3-constraints} and the hyperplane representation for generic $r$ (whose derivation is described in Sec.~\ref{sec:QMP}) constitute a generalization of the \textit{spin-adapted} Pauli exclusion principle constraints \eqref{eq:spin-Pauli} to $\bd w$-ensemble states.

\subsection{Singlet constraints\label{sec:singlet}}

The ground state of many atomic or molecular systems is a singlet state. We therefore derive in the following the constraints for $S=0$ and generic $N, d$ for $r\leq 3$. The $\wb$-ensemble exclusion principle constraints for arbitrary $r$ can again be calculated by following the general derivation of the vertex and hyperplane representation in Sec.~\ref{sec:QMP}. As indicated above, the case $S=0$ requires particular care since the stability constraint \eqref{eq:stability3} is only met for $r=1$. In turn this means that we have to study the distinctive excitation diagrams for $S=0$, such as the one shown in Fig.~\ref{fig:singlet-spectrum}.

\begin{figure*}[htb]
\includegraphics[width=\linewidth]{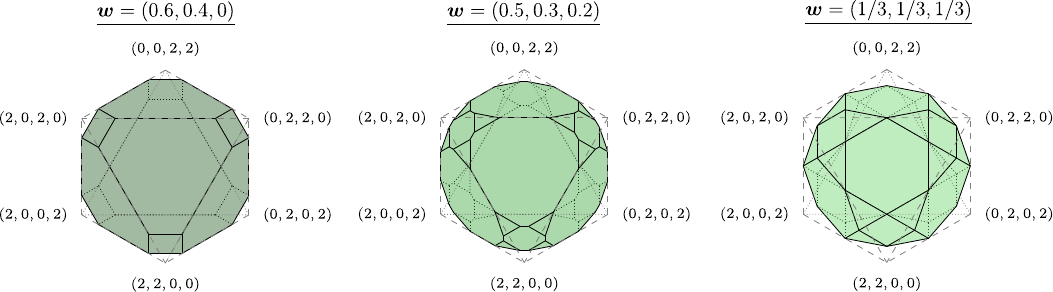}
\caption{Illustration of the spectral polytope $\Sigma_{N,S}(\bd w)$ for $N=d=4, S=0$ and different vectors of $\bd w$ for $r=2,3$ non-vanishing weights in $\wb$. The spectral set $\Sigma_{N,S}(1,0,\ldots)$ that corresponds to the Pauli constraints for the chosen $S, N, d$ is shown in gray. The left panel corresponds to $r=2$ non-zero weights and, thus there is only one additional $\bd w$-dependent inequality \eqref{eq:r2-singlet-constraint} that cuts out the corners of the Pauli polytope. For $r=3$ (middle and right panel), this constraint is still facet-defining but there is one new constraint resulting in additional facets. \label{fig:poly_S0}}
\end{figure*}

We first recall from Sec.~\ref{sec:Hasse} that for $S=0$ the spectrum of the orbital 1RDM corresponding to the unique highest weight vector is given by
\begin{equation}
\bd \lambda^{(\Lambda)} =\mathrm{spec}^\downarrow(\mu(\ket{\Lambda}\!\bra{\Lambda})) = (2, 2, \ldots, 2, 0, \ldots)\,.
\end{equation}
The Hasse diagram of partially ordered configurations for generic $N,d$ and $S=0$ is constructed as described in Sec.~\ref{sec:Hasse} and illustrated as an example in Fig.~\ref{fig:singlet-spectrum}.
In particular, we obtain from Sec.~\ref{sec:Hasse} that the unique lineup for $r=2$ is given by
\begin{align}
&\left(1,1, \ldots, \frac{N}{2},\frac{N}{2}\right)\nonumber\\
&\quad\to\left(1, 1, \ldots, \frac{N}{2}-1, \frac{N}{2}-1, \frac{N}{2}, \frac{N}{2}+1\right)\,.
\end{align}
According to Eq.~\eqref{eq:vertex}, this yields the generating vertex
\begin{equation}
\bd v = (\underbrace{2, \ldots, 2}_{N/2-1}, 1+w_1, 1-w_1, 0, \ldots)\,.
\end{equation}
Thus, the hyperplane representation of the spectral polytope $\Sigma_{N,S}(\bd w)$ for $r=2$ follows directly from Rado's theorem. $\Sigma_{N,S}(\bd w)$ is characterized by the spin-adapted Pauli constraint following from \eqref{eq:spin-Pauli} for $S=0$ (see also Ref.~\cite{LMS23}), i.e.,
\begin{equation}\label{eq:spin-Pauli-sing}
\lambda_1^\downarrow \leq  2
\end{equation}
and the additional $\bd w$-dependent constraint
\begin{equation}\label{eq:r2-singlet-constraint}
\sum_{i=1}^{N/2}\lambda_i^\downarrow \leq N-1+w_1\,.
\end{equation}
The spectral polytope for $r=2$ and $N=d=4, \bd w = (0.6, 0.4, 0, \ldots)$ is illustrated in the left panel of Fig.~\ref{fig:poly_S0}. The forth entry of $\bd\lambda$ is discarded due to the normalization $\sum_{i=1}^4\lambda_i=4$, in analogy to the visualization in Fig.~\ref{fig:poly_S1}. The facet-defining hyperplane that removes the vertices of the Pauli simplex (gray) is determined by Eq.~\eqref{eq:r2-singlet-constraint}.

For $r=3$, there are three possible options for the third excitation as illustrated for the example of $N=6$ in Fig.~\ref{fig:singlet-spectrum}. Their explicit form for generic $N, d$ in provided in Appendix \ref{app:singlet}. These three lineups then yield the three generating vertices
\begin{align}\label{eq:S0-vertices-r3}
\bd v^{(1)}&= (\underbrace{2, \ldots, 2}_{\frac{N}{2}-1}, 1+w_1, w_2, 1-w_1-w_2, 0, \ldots)\,,\\
\bd v^{(2)}&= (\underbrace{2, \ldots, 2}_{\frac{N}{2}-1}, 2w_1+w_2, 2-2w_1-w_2, 0, \ldots)\,,\nonumber\\
\bd v^{(3)}&=(\underbrace{2, \ldots, 2}_{\frac{N}{2}-2}, 1+w_1+w_2, 2-w_2, 1-w_1, 0, \ldots)\,.\nonumber
\end{align}
We then show in Appendix \ref{app:singlet} that these three vertices yield in addition to the Pauli exclusion principle \eqref{eq:spin-Pauli-sing} and $\sum_{i=1}^N\lambda_i^\downarrow=N$ the following two constraints,
\begin{align}\label{eq:r3-singlet-constraints}
\sum_{i=1}^{N/2}\lambda_i^\downarrow &\leq N-1+w_1\,,\nonumber\\
3\sum_{i=1}^{N/2-1}\lambda_i^\downarrow+2\lambda_{N/2}^\downarrow+\lambda_{N/2+1}^\downarrow&\leq 3N-4+2w_1+w_2\,.
\end{align}
Thus, there exists one additional inequality, namely the second one in Eq.~\eqref{eq:r3-singlet-constraints}, in addition to the constraint for $r=2$ in Eq.~\eqref{eq:r2-singlet-constraint}. This demonstrates again the hierarchy in $r$ of the $\wb$-ensemble exclusion principle constraints. We present a graphical illustration of the polytopes $\Sigma_{N,S}(\bd w)$ in Fig.~\ref{fig:poly_S0}  for $N=d=4, S=0$. Comparing the three different panels clearly showcases the hierarchy in $r$ and confirms that the polytope's volume shrinks as the degree of mixedness of $\bd w$ increases (referring to majorization \eqref{eq:majoriz}), in analogy to Fig.~\ref{fig:poly_S1} for triplets.

For $r=4$, there are in general eight lineups, as discussed in Appendix \ref{app:singlet}, which yield the following eight vertices
\begin{widetext}
\begin{align}\label{eq:S0-vertices-r4}
\bd v^{(1)} &= (\underbrace{2, \ldots, 2}_{\frac{N}{2}-1},1+w_1,w_2,w_3,1-w_1-w_2-w_3, 0, \ldots)\,, \nonumber\\
\bd v^{(2)} &= (\underbrace{2, \ldots, 2}_{\frac{N}{2}-1},2w_1+w_2+w_3,2-2w_1-w_2-2w_3,w_3,0,\ldots)\,,  \nonumber\\
\bd v^{(3)} &= (\underbrace{2, \ldots, 2}_{\frac{N}{2}-1},1+w_1-w_3,w_2+2 w_3, 1-w_1-w_2-w_3,0, \ldots)\,, \nonumber\\
\bd v^{(4)} &= (\underbrace{2, \ldots, 2}_{\frac{N}{2}-2}, 1+w_1+w_2+w_3,2-w_2-2w_3,1-w_1+w_3,0,\ldots)\,, \nonumber\\
\bd v^{(5)} &= (\underbrace{2, \ldots, 2}_{\frac{N}{2}-2},2-w_3,2w_1+w_2+2w_3,2-2w_1-w-2-w_3,0,\ldots)\,,  \nonumber\\
\bd v^{(6)} &= (\underbrace{2, \ldots, 2}_{\frac{N}{2}-2},1+w_1+w_2+w_3,2-w_2-w_3,1-w_1-w_3,w_3,0,\ldots)\,, \nonumber\\
\bd v^{(7)} &= (\underbrace{2, \ldots, 2}_{\frac{N}{2}-2}, 2-w_3,1+w_1+w_3,w_2+w_3,1-w_1-w_2-w_3, 0, \ldots)\,,  \nonumber\\
\bd v^{(8)} &= (\underbrace{2, \ldots, 2}_{\frac{N}{2}-3},1+w_1+w_2+w_3,2-w_3,2-w_2, 1-w_1, 0, \ldots)\,.
\end{align}
\end{widetext}
These generating vertices yield the vertex representation of $\Sigma_{N,S}(\bd w)$ in virtue of Eq.~\eqref{eq:Sigma-vrep}.

We then transform this vertex representation into a minimal hyperplane representation (see Appendix \ref{app:singlet}). This eventually yields for $r=4$ and $S=0$ three additional $\bd w$-dependent constraints,
\begin{widetext}
\begin{align}\label{eq:r4-singlet-constraints}
3\sum_{i=1}^{N/2-1}\lambda_i^\downarrow+ 2\lambda_{N/2}^\downarrow+\lambda_{N/2+1}^\downarrow+\lambda_{N/2+2}^\downarrow&\leq 3N-4+2 w_1+w_2+w_3 \,,\nonumber\\
2\sum_{i=1}^{N/2-1}\lambda_i^\downarrow+\lambda_{N/2}^\downarrow+\lambda_{N/2+1}^\downarrow&\leq 2N-3+w_1+w_2+w_3 \,,\nonumber\\
3\sum_{i=1}^{N/2-2}\lambda_i^\downarrow+ 2\lambda_{N/2-1}^\downarrow+2\lambda_{N/2}^\downarrow+\lambda_{N/2+1}^\downarrow&\leq 3N-6+2w_1+w_2+w_3\,,
\end{align}
\end{widetext}
which complement the constraints for $r=3$, i.e., the Pauli exclusion principle \eqref{eq:spin-Pauli-sing} and the constraints in Eq.~\eqref{eq:r3-singlet-constraints}.

\section{Applications\label{sec:appl}}

In this section, we discuss three direct applications of the spin-adapted orbital one-body $\bd w$-ensemble $N$-representability constraints derived in the previous sections. They are all concerned with the task of calculating the lowest few energy eigenstates of an interacting $N$-electron system in virtue of the ensemble variational principle due to Gross, Oliveira and Kohn \cite{GOK88a} (see also Ref.~\cite{LHS24}): Let $H$ be a Hamiltonian with increasingly ordered eigenvalues $E_1\leq E_2 \leq \ldots \leq E_D$ on a $D$-dimensional Hilbert space $\mathcal{H}$ and denote the set of density matrices on $\mathcal{H}$ with spectrum $\wb=\wb^{\downarrow}$ by $\mathcal{E}(\bd w)$. Then,
\begin{eqnarray}\label{eq:GOK}
E_{\bd w} \equiv \sum_{i=1}^D w_i E_i &=& \min_{\Gamma\in \mathcal{E}(\bd w)}\Tr\left[H\Gamma\right]\nonumber\\
& =& \min_{\Gamma\in \overline{\mathcal{E}}(\bd w)  }\Tr\left[H\Gamma\right]\,,
\end{eqnarray}
i.e., the $\wb$-averaged energy $E_{\bd w}$ can be obtained \emph{variationally} by minimizing the energy expectation value $\Tr\left[H\Gamma\right]$ over all density matrices with spectrum $\wb$. The resulting minimum is not changed if one extends the variational principle to the convex hull $\overline{\mathcal{E}}(\bd w)$ of $\mathcal{E}(\bd w)$, i.e., by including also those density matrices $\Gamma$ whose spectrum is majorized by $\wb$ (recall the comment above Eq.~\eqref{eq:majoriz}).

\subsection{Functional theory for excited states\label{sec:rdmft}}

The scope of a one-body reduced density matrix functional theory (RDMFT) is characterized by a family of Hamiltonians of interest \cite{LCS23} that take the form $H(h_l)=h_l+W$ which we assume to commute with $\bd S^2, S_z$. Here, the interaction $W$ between the electrons (or any other spin-$1/2$ fermions) is kept fixed, while the orbital one-particle Hamiltonian $h_l$ on $\mathcal{H}_1^{(l)}$ can be varied \footnote{For the sake of simplicity, we use here the same symbol for that Hamiltonian as an operator on $\mathcal{H}_1^{(l)}$ and $\mathcal{H}_N$.}.
Moreover, we denote for a fixed choice of $S,M$ the increasingly ordered eigenvalues of $H(h_l)$ within $\mathcal{H}_N^{(S,M)}$ by $E_1^{(S,M)}(h_l)\leq E_2^{(S,M)}(h_l)\leq \ldots\leq E_D^{(S,M)}(h_l)$ with $D \equiv D_N^{(S,M)}$. Applying now the ensemble variational principle \eqref{eq:GOK} yields the weighted sum of the lowest $r$ eigenenergies within $\mathcal{H}_N^{(S,M)}$,
\begin{eqnarray}
E_{\bd w}^{(S,M)}(h_l) &\equiv& \sum_{i=1}^rw_i E_i^{(S,M)} \nonumber \\
 &=& \min_{\Gamma\in \mathcal{E}^N_{S,M}(\bd w)}\Tr_N\left[H(h_l)\Gamma\right]\nonumber\\
& =& \min_{\Gamma\in \EbwS}\Tr_N\left[H(h_l)\Gamma\right]\,.
\end{eqnarray}
Moreover, it follows from the ensemble variational principle that $E_{\bd w}^{(S,M)}(h_l)$ is a concave functional of $h_l$.
Accordingly, we obtain a spin-adapted $\bd w$-ensemble RDMFT directly by applying the ideas in Refs.~\cite{SP21, LCLS21, LS23-sp, LS23-njp} to each spin sector $\mathcal{H}_N^{(S,M)}$ separately. The universal interaction functional follows from the Legendre-Fenchel transformation of $E_{\bd w}^{(S,M)}(h_l)$ in analogy to Lieb's convex formulation of DFT \cite{L83, HT22} according to
\begin{equation}\label{eq:FSM-LF}
\xxbar{\mathcal{F}}^{(S,M)}(\bd w)\equiv \sup_{h_l}\left(E_{\bd w}^{(S,M)}(h_l) - \Tr_{\mathcal{H}_1^{(l)}}[h_l \gamma_l]  \right)\,.
\end{equation}
Therefore, the functional $\xxbar{\mathcal{F}}^{(S,M)}(\bd w)$ is convex by definition and thus equal to the  ensemble constrained search functional \cite{V80}
\begin{equation}\label{eq:FSM-V}
\xxbar{\mathcal{F}}^{(S,M)}(\bd w)=\min_{\EbwS\ni\Gamma\mapsto\gamma_l}\Tr_N[\Gamma W]\,.
\end{equation}
Moreover, following Ref.~\cite{Schilling18} reveals that $\xxbar{\mathcal{F}}^{(S,M)}(\bd w)$ is the lower convex hull, $\xxbar{\mathcal{F}}^{(S,M)}(\bd w) = \mathrm{conv}(\mathcal{F}^{(S,M)}(\bd w))$, of the (typically non-convex) pure constrained search functional \cite{Levy79},
\begin{equation}
\mathcal{F}^{(S,M)}(\bd w) \equiv \min_{\mathcal{E}^N_{S,M}(\bd w)\ni\Gamma\mapsto\gamma_l}\Tr_N[\Gamma W]\,.
\end{equation}
The functionals $\mathcal{F}^{(S,M)}(\bd w)$ and $\xxbar{\mathcal{F}}^{(S,M)}(\bd w)$ both depend explicitly on the quantum numbers $S,M$ as well as the weight vector $\bd w$. More importantly, their domains are given precisely by the set $\mathcal{L}_{N,S}^1(\wb)$ and its convex hull $\ebw$, respectively. The latter is nothing else than the set of orbital 1RDMs that are compatible to an $N$-fermion quantum state with well-defined spin quantum numbers $S,M$ and spectrum majorized by the vector $\bd w$. With the compact characterization of that domain at hand derived in Secs.~\ref{sec:QMP}, \ref{sec:examples}, the process of deriving more and more accurate functional approximations for $\xxbar{\mathcal{F}}^{(S,M)}(\bd w)$ can now commence.

\subsection{Lattice density functional theory\label{sec:lattice-DFT}}

The application of the solution to the orbital one-body $\bd w$-ensemble $(N,S,M)$-representability problem in functional theories is not limited to RDMFT. In this section, we demonstrate that it also constrains the set of admissible occupation number vectors in the so-called ensemble DFT for excited states (EDFT) based on the ensemble variational principle \eqref{eq:GOK} \cite{Theophilou79, GOK88c, TG95, YPBU17, Fromager2020-DD, Loos2020-EDFA, Cernatic21, Yang21, GKGGP23, SKCPJB24, CPSF24} when applied to lattice systems with a global $SU(2)$ spin symmetry.
For an introduction to lattice DFT, see Refs.~\cite{GS86, SGN95, IH10, XCTK12, SP14, Coe19, PvL21, PvL23, PvL24}. In DFT, we consider Hamiltonians $H(v)=v+t+W$, which are parameterized by a local external potential $v$ while the kinetic energy operator $t$ and the interaction $W$ are kept fixed.
Then, the natural variable in lattice DFT is the vector $\bd \rho\in\RR^d$ of (site) occupation numbers $\rho_i = (\gamma_l)_{ii}$, which are the conjugate variables to the local external potentials $v_i$ on the lattice sites $i\in\{1, \ldots, d\}$. Through the dual pairing, the expectation value of the local external potential follows as $\langle\bd v, \bd\rho\rangle = \sum_{i=1}^d v_i\rho_i$.
Excited states can be calculated by means of the ensemble variational principle, as explained in detail in the review article \cite{Cernatic21}. Similarly to Eqs.~\eqref{eq:FSM-LF} and \eqref{eq:FSM-V}, the ensemble universal functional in lattice GOK-DFT is defined as
\begin{equation}
\xxbar{\mathcal{G}}^{(S,M)}(\bd \rho)\equiv \min_{\EbwS\ni\Gamma\mapsto\bd\rho}\Tr_N[(t+W)\Gamma]\,.
\end{equation}
The domain of $\xxbar{\mathcal{G}}^{(S,M)}(\bd \rho)$ is given by all those lattice site occupation number vectors $\bd\rho$, which are compatible to an $N$-particle state $\Gamma\in \Ebw$ and, thus, also to an orbital 1RDM $\gamma_l\in \ebw$. The domain of $\xxbar{\mathcal{G}}^{(S,M)}(\bd \rho)$ explicitly depends on the total spin quantum number $S$ as well as the weight vector $\bd w$, as shown in the following.

To this end, we first recall from Ref.~\cite{LCLS21} that a vector $\bd \lambda$ is an element of the permutation invariant polytope $\Sigma_{N,S}(\bd w)$ if and only if it is majorized by a convex combination $\bd u \equiv \sum_{i=1}^R q_i\bd v^{(i)}$ of the generating vertices $\bd v^{(i)}, i\in\{1, \ldots, R\}$, that is
\begin{equation}
\bd\lambda\in\Sigma_{N,S}(\bd w)\Leftrightarrow \bd\lambda\prec\bd u \equiv \sum_{i=1}^R q_i\bd v^{(i)},\,\,q_i\geq 0, \sum_{i=1}^R q_i=1\,.
\end{equation}
Moreover, $\bd \rho$ is majorized by the vector $\bd\lambda$ of eigenvalues of $\gamma_l$ as a consequence of the Schur-Horn theorem \cite{Schur1923, Horn54}. Thus, it follows from the transitivity of the majorization that
\begin{equation}
\bd\rho\prec \bd\lambda\prec \bd u\,.
\end{equation}
We therefore conclude that the domain of $\xxbar{\mathcal{G}}^{(S,M)}(\bd \rho)$ is determined precisely by the relaxed orbital one-body $\bd w$-ensemble $N$-representability constraints derived in Secs.~\ref{sec:QMP}, \ref{sec:examples}, yet here applied to $\bd \rho$ rather than $\bd \lambda$. Besides its $\bd w$-dependence, this also shows that the domain of $\xxbar{\mathcal{G}}^{(S,M)}(\bd \rho)$ explicitly depends on the total spin quantum number $S$.

\subsection{Contraction conditions for higher order reduced density matrices \label{sec:prdms}}

As a third application, the orbital one-body $\bd w$-ensemble $(N,S,M)$-representability constraints provide necessary but not sufficient conditions for the $\bd w$-ensemble $(N,S,M)$-representability of higher-order reduced density matrices (RDMs). This includes the full 1RDM $\gamma = N\Tr_{N-1}[\Gamma], \Gamma\in \Ebw$, as well as $p$-particle reduced density matrices ($p$RDMs) with $p\geq 2$.

We denote by
\begin{equation}
\xxbar{\mathcal{E}}^{p}_{N,S,M}(\bd w)\equiv \begin{pmatrix}
N\\ p
\end{pmatrix}\Tr_{N-p}\left[\Ebw\right]
\end{equation}
the set of $p$RDMs $\Gamma^{(p)}$ compatible with an $N$-fermion state $\Gamma\in\Ebw$. The restriction to $N$-fermion states $\Gamma\in\Ebw$ instead of all $N$-fermion density operators imposes additional $\bd w$-dependent constraints beyond the previously derived ensemble $p$-body $N$-representability conditions \cite{C63, GarPer64, Kummer67, M12-prl, M16-pra, M23-prl}. Every $\Gamma^{(p)}\in \xxbar{\mathcal{E}}^{p}_{N,S,M}(\bd w)$ satisfies
\begin{align}\label{eq:contraction}
\mu^{(p)}\left(\Gamma^{(p)}\right) \in \ebw\,,
\end{align}
where we defined the linear function $\mu^{(p)}$ which maps a $p$RDM to its orbital 1RDM according to
\begin{align}
\mu^{(p)}:\xxbar{\mathcal{E}}^{p}_{N,S,M}(\bd w)&\to \ebw\nonumber\\
\Gamma^{(p)} &\mapsto\gamma_l\,,
\end{align}
where $\mathcal{H}_p \equiv \wedge^p\mathcal{H}_1$ denotes the Hilbert space of $p$ spin-$1/2$ fermions.

Solving the relaxed $\bd w$-ensemble $(N,S,M)$-representability problem for $p$RDMs is highly complex even for $p=1$, since the corresponding sets $\xxbar{\mathcal{E}}^p_{N,S,M}(\bd w)$ are not unitarily invariant anymore. This in turn highlights the potential significance of the contraction relation \eqref{eq:contraction}. For instance, it might be valuable for implementing variational 2RDM approaches \cite{M04, M05, CL06, VAA12, DLBBKRB23, M20, DP24} within the $\bd w$-ensemble framework based on the ensemble variational principle \cite{GOK88a,LHS24}.

\section{Summary and outlook\label{sec:summary}}

Motivated by recent advances in reduced density matrix methods, we have developed a comprehensive solution to the convex-relaxed, spin-adapted one-body $\bd{w}$-ensemble $N$-representability problem for the orbital 1RDM $\gamma_l$. We demonstrated that the set of admissible $\gamma_l$ is defined by linear spectral constraints on the natural orbital occupation numbers $\lambda_i$, forming a convex polytope $\Sigma_{N,S}(\bd{w}) \subset [0,2]^d$. These $\bd{w}$-ensemble exclusion principle constraints refine the Pauli exclusion principle, $0 \leq \lambda_i \leq 2$, and depend linearly on $N$ and $S$, while being independent of $M$ and the number $d$ of orbitals. This allows their computation for arbitrary spin quantum numbers and system sizes, including the infinite basis set limit. Additionally, we uncovered a hierarchical structure in these constraints as a function of the number $r$ of non-vanishing entries in $\bd{w}$: constraints for $r$ are given by those for $r-1$, supplemented by new ones. Importantly, our constraints explicitly account for spin, a feature essential for many quantum physics and chemistry applications, distinguishing them from prior works.

We also demonstrated three applications of the spin-dependent $\bd{w}$-ensemble constraints in reduced density matrix methods (Sec.~\ref{sec:appl}). First, these constraints are pivotal for $\bd{w}$-RDMFT in fermionic systems with global $SU(2)$ spin symmetry, enabling the calculation of low-lying excited states in specific spin sectors. Through the constrained search formalism, we establish the universal interaction functional of the orbital 1RDM $\gamma_l$, whose domain is precisely described by these constraints. Second, as a consequence of the Schur-Horn theorem \cite{Schur1923, Horn54}, the constraints also define the set of admissible densities in ensemble density functional theory (EDFT). Third, they provide a first outer approximation to the set of $\bd{w}$-ensemble $(N, S, M)$-representable $p$-RDMs, with potential significance for variational 2RDM approaches \cite{M04, M05, CL06, VAA12, DLBBKRB23, M20, DP24} within the $\bd{w}$-ensemble framework.

Beyond RDM theories, the vertices of the spectral polytopes $\Sigma_{N,S}(\bd{w})$ correspond to spin-symmetry restricted Hartree-Fock (HF) states for $\bd{w}$-ensembles. By minimizing the energy expectation value over these states, we establish a mean-field theory for targeting excited states. The distance of a given $N$-electron density matrix's natural orbital occupation numbers $\bd{\lambda}$ from the vertices of $\Sigma_{N,S}(\bd{w})$ serves as both a metric for the accuracy of spin-adapted $\bd{w}$-HF calculations and a measure of correlation beyond spin-adapted HF states. This insight paves the way for a theory of electron correlation that quantifies quantum correlations beyond spin- and exchange-symmetry, extending to mixed states.

\begin{acknowledgments}
We acknowledge financial support from the German Research Foundation (Grant SCHI 1476/1-1) (J.L., C.S.) and the International Max Planck Research School for Quantum Science and Technology (IMPRS-QST) (J.L.). The project/research is also part of the Munich Quantum Valley, which is supported by the Bavarian state government with funds from the Hightech Agenda Bayern Plus.
\end{acknowledgments}

\appendix

\section{Derivation of relaxed orbital one-body $\bd w$-ensemble $N$-representability constraints}

\subsection{Generic $N,S,d$\label{app:genS}}

We derive the relaxed orbital one-body $\bd w$-ensemble $N$-representability constraints for $N, S, d$ satisfying Eqs.~\eqref{eq:stability1}-\eqref{eq:stability3} for $r\leq 3$. The constraints for any larger $r$ can be derived in an analogous manner. In the derivation we exploit the so-called hierarchy of exclusion principle constraints explained in the Sec.~\eqref{sec:examples}, which states that the constraints for $r+1$ consist of the constraints for $r$ and a finite number of additional ones. We therefore start with $r=2$.

For generic $N,S,d$ and $r=2$, we obtain (recall also the example in Fig.~\ref{fig:spectrum-generic}) two distinct lineups,
\begin{widetext}
\begin{align}\label{eq:lineups-r2}
&l_1\!: \underbrace{\left(1,1,2,2,\ldots, K, K, K+1, \ldots,  J\right)}_{\quad \,\,\,\equiv \, \bd i_1^{(1)}} \to\underbrace{\left(1,1,2,2,\ldots, K,K, K+1, \ldots,  J-1, J+1\right)}_{\quad \,\,\,\equiv \,\bd i_2^{(1)}}\,, \nonumber\\
&l_2\!: \underbrace{\left(1,1,2,2,\ldots, K, K, K+1, \ldots,  j\right)}_{\quad \,\,\,\equiv \, \bd i_1^{(2)}}\to\underbrace{\left(1,1,2,2,\ldots, K-1,K-1, K, K+1, K+1,K+2,\ldots,  J\right)}_{\quad \,\,\,\equiv \,\bd i_2^{(2)}}\,,
\end{align}
\end{widetext}
where $K=(N-2S)/2, J=(N+2S)/2$.
These two lineups correspond, in full analogy to Refs.~\cite{SP21, LCLS21}, to two distinctive $N$-fermion states, namely,
\begin{align}\label{eq:states-r2}
\Gamma_{\bd w}^{(1)} &= w\ket{\bd i_1^{(1)}}\!\bra{\bd i_1^{(1)}}+(1-w)\ket{\bd i_2^{(1)}}\!\bra{\bd i_2^{(1)}}\nonumber\\
\Gamma_{\bd w}^{(2)} &= w\ket{\bd i_1^{(2)}}\!\bra{\bd i_1^{(2)}}+(1-w)\ket{\bd i_2^{(2)}}\!\bra{\bd i_2^{(2)}}\,.
\end{align}
For both states $\Gamma_{\bd w}^{(1)}, \Gamma_{\bd w}^{(2)}$ we calculate the orbital 1RDM $\gamma^{(1/2)}_l$ via the map $\mu(\cdot)$ defined in Eq.~\eqref{eq:map}. Then, the generating vertices $\bd v^{(1/2)}$ of the spectral polytope $\Sigma_{N,S}(\bd w)$ follow as the spectra of $\gamma_l^{(1/2)}$ according to
\begin{equation}
\bd v^{(i)} = \mathrm{spec}\left(\gamma^{(i)}_l\right)\,.
\end{equation}
For the two states $\Gamma_{\bd w}^{(1)}, \Gamma_{\bd w}^{(2)}$ in Eq.~\eqref{eq:states-r2}, we eventually obtain the two generating vertices in Eq.~\eqref{eq:v-r2}.

In the next step, we apply the formalism outlined in Sec.~\ref{sec:H-repr} to translate the vertex representation of $\Sigma_{N,S}(\bd w)$ into a minimal hyperplane representation.
By using the two lineups in Eq.~\eqref{eq:lineups-r2}, we first determine the normal cones of the vertices of $\Sigma_{N,S}(\bd w)$. To this end, we have calculate the set of linear functionals, which are uniquely maximized or, equivalently, minimized at a generating vertex $\bd v^{(i)}$. We say that a linear functional $\bd\eta$ enforces a lineup $l_i$ if it is contained in the (open) normal cone of the vertex $\bd v^{(i)}$.
To enforce the first lineup $l_1$, the fundamental linear functional $\bd\eta$ has to satisfy (recall Eq.~\eqref{eq:kj})
\begin{equation}\label{eq:tmp1}
\langle\bd\eta, \bd e_K+\bd e_{J+1}-\bd e_{K+1} - \bd e_J\rangle >0\,,
\end{equation}
where $\bd e_i= (0, \ldots, 1, 0, \ldots)$ are the basis vectors of the standard basis. Thus, all entries of $\bd e_i$ are zero except for a single `$1$' in the $i$-th entry. Moreover, using $\bd f_i$ defined in Eq.~\eqref{eq:f-dual} we see that the fundamental linear functional $\bd\eta$ is given by
\begin{equation}
\bd\eta = \sum_{i=1}^d\eta_i\bd f_i
\end{equation}
with $\eta_i\geq 0\,\,\forall i=1, .., d-1$ and $\eta_d\in \RR$. Then, Eq.~\eqref{eq:tmp1} simplifies to
\begin{equation}
\eta_{K}-\eta_J >0\,.
\end{equation}
Similarly, we obtain that the second lineup $l_2$ is enforced if $\eta_{K}-\eta_J <0$. The two inequalities imply the ray $\bd f_K + \bd f_J$ as explained in Sec.~\ref{sec:H-repr}. The left hand-side of the corresponding inequality then follows from taking the inner product of all rays with $\bd\lambda^\downarrow$ and the right hand-side from evaluating the rays at the generating vertices. This leads to a non-trivial $\bd w$-dependent linear constraint on $\bd \lambda$ given by
\begin{equation}\label{eq:r2-w}
\langle \bd f_K+\bd f_J, \bd\lambda^\downarrow\rangle = 2\sum_{i=1}^K \lambda_i^\downarrow+\sum_{j=K+1}^J\lambda_j^\downarrow \leq 2(N-S)-1+w_1 \,.
\end{equation}
The hyperplane representation of $\Sigma_{N,S}(\bd w)$ for $r=2$ is then given by the constraint in Eq.~\eqref{eq:r2-w} and the spin-adapted Pauli constraints derived in Ref.~\cite{LMS23}.

Next, we derive the spin-dependent $\bd w$-ensemble exclusion principle constraints that characterize the hyperplane representation of $\Sigma_{N,S}(\bd w)$ for $r=3$.
The Hasse diagram of partially ordered configurations yields the following $R=6$ lineups
\begin{widetext}
\begin{eqnarray}\label{eq:lineups-r3}
&&l_1: (1, 1, \ldots, K,K, K+1, \ldots, J) \to (1, 1, \ldots, K-1, K-1, K, K+1, K+1, K+2, \ldots, J)\nonumber\\
&&\quad\quad\to (1, 1, \ldots, K,K,K+1, \ldots, J-1, J+1)\,, \\
&&l_2: (1, 1, \ldots, K,K, K+1, \ldots, J) \to (1, 1, \ldots, K,K,K+1, \ldots, J-1, J+1)\nonumber\\
&&\quad\quad\to (1, 1, \ldots, K-1, K-1, K, K+1, K+1, K+2, \ldots, J) \,,\nonumber\\
&&l_3: (1, 1, \ldots, K,K, K+1, \ldots, J) \to (1, 1, \ldots, K-1, K-1, K, K+1, K+1, K+2, \ldots, J)\nonumber\\
&&\quad\quad\to (1,1,\ldots, K-1, K-1, K, K+1, K+2, K+2, K+3, \ldots, J) \,,\nonumber\\
&&l_4: (1, 1, \ldots, K,K, K+1, \ldots, J) \to (1, 1, \ldots, K-1, K-1, K, K+1, K+1, K+2, \ldots, J)\nonumber\\
&&\quad\quad\to (1, 1, \ldots, K-2, K-2, K-1, K,K,K+1, K+1, K+2, \ldots, J) \,,  \nonumber\\
&&l_5:  (1, 1, \ldots, K,K, K+1, \ldots, J) \to (1, 1, \ldots, K,K,K+1, \ldots, J-1, J+1)\nonumber\\
&&\quad\quad\to (1, 1, \ldots, K,K,K+1, \ldots, J-1, J+2) \,, \nonumber\\
&&l_6:  (1, 1, \ldots, K,K, K+1, \ldots, J) \nonumber\\
&&\quad\quad\to (1, 1, \ldots, K,K,K+1, \ldots, J-1, J+1)\to (1,1,\ldots, K,K,K+1, \ldots, J-2, J, J+1)\,.\nonumber
\end{eqnarray}
\end{widetext}
From the lineups in Eq.~\eqref{eq:lineups-r3}, we eventually obtain the six generating vertices $\bd v^{(i)}, i=1, \ldots, 6$ shown in Eq.~\eqref{eq:v-r3} which fully characterize the vertex representation of $\Sigma_{N,S}(\bd w)$ in Eq.~\eqref{eq:Sigma-vrep} for $r=3$ non-zero weights $w_i$.

Next, we determine the conditions on the fundamental linear functional $\bd\eta$ such that they enforce the lineups $l_i, i=1, \ldots, 6$.
Let us start with the first lineup $l_1$. According to Sec.~\ref{sec:r2} the orbital configuration in its second entry is enforced by $\eta_J-\eta_K>0$. Moreover, there are three possible orbital configurations for the third position in the lineup having fixed the second one. To enforce the lineup $l_1$, the fundamental linear functionals $\bd\eta$ thus need to satisfy both
\begin{eqnarray}
\langle \bd\eta, \bd e_K+\bd e_{J+1}-\bd e_{K+2}-\bd e_J\rangle &>&0\,,\nonumber\\
\langle \bd\eta, \bd e_{J+1}+\bd e_{K-1}-\bd e_{K+1}-\bd e_J\rangle &>& 0\,,
\end{eqnarray}
which is equivalent to
\begin{eqnarray}
\eta_K+\eta_{K+1}>\eta_J &>&\eta_K\,,\nonumber \\
\eta_{K-1}+\eta_K> \eta_J &>&\eta_K\,.
\end{eqnarray}
Similarly to $l_1$, the second lineup $l_2$ is enforced if $\eta_K-\eta_J>0$ in combination with
\begin{eqnarray}
\eta_J+\eta_{J+1} &>& \eta_K >\eta_J
\end{eqnarray}
from enforcing $l_2$ over $l_5$ and
\begin{equation}
\eta_{J-1}+\eta_J>\eta_K>\eta_J
\end{equation}
is obtained from enforcing $l_2$ over $l_6$.
The second position in the third lineup $l_3$ is enforced by $\eta_J-\eta_K>0$. To enforce the correct third position in $l_3$, we furthermore require that
\begin{eqnarray}
\eta_J &>&\eta_K+\eta_{K+1}\,,\nonumber\\
\eta_{K-1}&>&\eta_{K+1}\,.
\end{eqnarray}
To enforce the second entry in the forth lineup $l_4$ we need $\eta_J-\eta_K>0$ and the third position requires
\begin{eqnarray}
\eta_{K+1}&>&\eta_{K-1}\,,\nonumber \\
\eta_J&>&\eta_{K-1}+\eta_K\,.
\end{eqnarray}
Moreover, the linear functional $\bd \eta$ enforces the lineup $l_5$ if $\eta_K-\eta_J>0$ and in addition
\begin{eqnarray}
\eta_K &>&\eta_J+\eta_{J+1}\,,\nonumber\\
\eta_{J-1}&>&\eta_{J+1}\,.
\end{eqnarray}
For the sixth lineup $l_6$, we finally obtain in addition to $\eta_K>\eta_J$ the constraint
\begin{eqnarray}
\eta_{J+1}&>&\eta_{J-1}\,,\nonumber\\
\eta_K&>&\eta_{J-1}+\eta_J\,.
\end{eqnarray}

As in Appendix \ref{app:genS} one of the resulting rays is $\bd f_K+\bd f_J$. From $l_1, l_3, l_4$ we further obtain the ray $\bd f_{K-1}+\bd f_{K+1}+\bd f_{J}$. In addition the constraints on $\bd\eta$ from $l_2, l_5, l_6$ generate the ray $\bd f_{K}+\bd f_{J-1}+\bd f_{J+1}$. These rays are complemented by $\bd f_K, \bd f_J$.
Evaluating these rays on the six vertices to obtain the right hand-side of the inequalities finally yields
\begin{align}
2\sum_{i=1}^{K}\lambda_i^\downarrow+\sum_{i=K+1}^{J}\lambda_i^\downarrow&\leq 2(N-S)-1+w_1 \,,\nonumber\\
3\sum_{i=1}^{K-1}\lambda_i^\downarrow + 2\sum_{i=K}^{K+1}\lambda_i^\downarrow+\sum_{i=K+2}^J\lambda_i^\downarrow &\leq 3N-4S-2+w_1+w_2\,,\nonumber\\
3\sum_{i=1}^{K}\lambda_i^\downarrow + 2\sum_{i=K+1}^{J-1}\lambda_i^\downarrow+\sum_{i=J}^{J+1}\lambda_i^\downarrow &\leq 3N-2S-2+w_1+w_2\,,\nonumber\\
\sum_{i=1}^K\lambda_i^\downarrow&\leq N-2S\,,\nonumber\\
\sum_{i=1}^J\lambda_i^\downarrow&\leq N\,.
\end{align}

\subsection{Singlet setting and $r=3,4$\label{app:singlet}}

In this section, we explicitly derive the relaxed orbital one-body $\bd w$-ensemble $N$-representability constraints for singlets, i.e., $S=0$, and $r=3,4$ non-zero weights $w_i$. The simple case of $r=2$ is discussed in the main text in Sec.~\ref{sec:singlet}. The singlet setting requires a separate treatment from the general case in Appendix \ref{app:genS} since the three stability conditions in Eqs.~\eqref{eq:stability1}-\eqref{eq:stability3} are not met for $S=0$. The derivation of the additional $\bd w$-dependent constraint for $r=2$ was shown in the main text and reduces to the spin-independent constraint for $r=2$ in Refs.~\cite{SP21, LCLS21, CLLPPS23}. For larger $r$ the constraints for $S=0$ do not coincide with the spin-independent constraints anymore, as a result of the different Hasse diagram of partially ordered configurations, which was illustrated for up to three excitations on top of the non-interacting ground state and $N=6$ in Fig.~\ref{fig:singlet-spectrum}.

For $r=3$, i.e., states with fixed spectra $\bd w=(w_1, w_2, 1-w_1-w_2, 0, \ldots)$ there are three lineups given by
\begin{widetext}
\begin{align}
&l_1:\,\,\left(1, 1, \ldots, \frac{N}{2}, \frac{N}{2}\right)\to \left(1, 1, \ldots, \frac{N}{2}-1, \frac{N}{2} -1, \frac{N}{2}, \frac{N}{2}+1\right)\to \left(1, 1, \ldots, \frac{N}{2}-1, \frac{N}{2}-1, \frac{N}{2}, \frac{N}{2}+2\right)\,,\nonumber\\
&l_2:\,\,\left(1, 1, \ldots, \frac{N}{2}, \frac{N}{2}\right)\to \left(1, 1, \ldots, \frac{N}{2}-1, \frac{N}{2} -1, \frac{N}{2}, \frac{N}{2}+1\right)\to \left(1, 1, \ldots, \frac{N}{2}-1, \frac{N}{2}-1, \frac{N}{2}+1, \frac{N}{2}+1\right)\,,\nonumber\\
&l_3:\,\,\left(1, 1, \ldots, \frac{N}{2}, \frac{N}{2}\right)\to \left(1, 1, \ldots, \frac{N}{2}-1, \frac{N}{2} -1, \frac{N}{2}, \frac{N}{2}+1\right)\to\left(1, 1, \ldots, \frac{N}{2}-2, \frac{N}{2}-2, \frac{N}{2}-1, \frac{N}{2}, \frac{N}{2}, \frac{N}{2}+1\right)\,.
\end{align}
\end{widetext}
As explained in Sec.~\ref{sec:v-repr}, they yield the three generating vertices in Eq.~\eqref{eq:S0-vertices-r4}.
Since the first two entries in the three lineups are equal, we only have to investigate the third entry to derive the hyperplane representation of $\Sigma_{N,0}(\bd w)$.
The first lineup is induced unique if a fundamental linear functional $\bd\eta$ satisfies
\begin{align}\label{eq:S0-r3-l1}
\eta_{\frac{N}{2}}&>\eta_{\frac{N}{2}+1}
\end{align}
and in addition
\begin{align}
\eta_{\frac{N}{2}-1}&>\eta_{\frac{N}{2}+1}\,.
\end{align}
To determine the two strict inequalities that induce $l_2$ uniquely, we first have to invert Eq.~\eqref{eq:S0-r3-l1} which yields
\begin{equation}
\eta_{\frac{N}{2}+1}>\eta_{\frac{N}{2}}
\end{equation}
and, second, to ensure that $l_2$ is obtained instead of $l_3$, $\bd\eta$ has to satisfy
\begin{align}
\eta_{\frac{N}{2}-1}&>\eta_{\frac{N}{2}}\,.
\end{align}
It follows that the third lineup is induced uniquely if
\begin{align}
\eta_{\frac{N}{2}+1}&>\eta_{\frac{N}{2}-1}\,,\nonumber \\
\eta_{\frac{N}{2}}&>\eta_{\frac{N}{2}-1}\,.
\end{align}
The inequalities lead to two rays, the ray $\bd f_{N/2}$ obtained for $r=2$ and a new ray $\bd f_{N/2-1}+\bd f_{N/2}+ \bd f_{N/2+1}$ which leads to an additional $\bd w$-dependent exclusion principle constraint. Thus, the hyperplane representation of $\Sigma_{N,0}(\bd w)$ for $r=3$ is again given by the spin-adapted Pauli constraints \cite{LMS23} and the now two new $\bd w$-dependent constraints
\begin{align}\label{eq:r3-singlet}
\sum_{i=1}^{N/2}\lambda_i^\downarrow &\leq N-1+w_1\,,\nonumber\\
3\sum_{i=1}^{N/2-1}\lambda_i^\downarrow+2\lambda_{N/2}^\downarrow+\lambda_{N/2+1}^\downarrow&\leq 3N-4+2w_1+w_2\,.
\end{align}

For $r=4$, there are in general eight lineups that yield the eight vertices in Eq.~\eqref{eq:S0-vertices-r4} in the main text in Sec.~\ref{sec:singlet}.
In analogy to the previous cases we then determine the corresponding rays of $\Sigma_{N,0}(\bd w)$ that follow from the eight lineups and the vertex representation of the polytope. Besides the trivial rays from the spin-adapted Pauli constraints, there are the two rays $\bd f_{N/2}, \bd f_{N/2-1}+\bd f_{N/2}+ \bd f_{N/2+1}$ from $r=3$ and the three new rays $\bd f_{N/2-1}+\bd f_{N/2}+\bd f_{N/2+2}, \bd f_{N/2-1}+\bd f_{N/2+1},\bd f_{N/2-2}+\bd f_{N/2}+\bd f_{N/2+1}$. The first two rays lead to the two inequalities in Eq.~\eqref{eq:r3-singlet}, while the three new rays yield the additional constraints
\begin{widetext}
\begin{align}
3\sum_{i=1}^{N/2-1}\lambda_i^\downarrow+ 2\lambda_{N/2}^\downarrow+\lambda_{N/2+1}^\downarrow+\lambda_{N/2+2}^\downarrow&\leq 3N-4+2 w_1+w_2+w_3 \,,\nonumber\\
2\sum_{i=1}^{N/2-1}\lambda_i^\downarrow+\lambda_{N/2}^\downarrow+\lambda_{N/2+1}^\downarrow&\leq 2N-3+w_1+w_2+w_3 \,,\nonumber\\
3\sum_{i=1}^{N/2-2}\lambda_i^\downarrow+ 2\lambda_{N/2-1}^\downarrow+2\lambda_{N/2}^\downarrow+\lambda_{N/2+1}^\downarrow&\leq 3N-6+2w_1+w_2+w_3\,.
\end{align}
\end{widetext}

\bibliographystyle{quantum}
\bibliography{Refs}

\end{document}